\documentclass[a4paper,12pt]{article}
\pdfoutput=1 

\usepackage{jheppub} 
\usepackage{hyperref}
\usepackage{subfigure}
\usepackage[T1]{fontenc} 
\usepackage{txfonts}
\usepackage{verbatim}
\usepackage{url}

\usepackage{booktabs}
\usepackage{siunitx}
\newcolumntype{d}{S[
	input-open-uncertainty=,
	input-close-uncertainty=,
	parse-numbers = false,
	table-align-text-pre=false,
	table-align-text-post=false
	]}

\usepackage{setspace}

\definecolor{mypink1}{rgb}{0.458, 0.218, 0.178}
\usepackage[utf8]{inputenc} 

\title{\fontfamily{ptm}\selectfont Measuring the impact of green retrofits on $CO_2$ emissions: London Housing Case Study}

\author[]{F. Correia,}
\author[]{M. Fazion}

\emailAdd{fagner.correia@susy.house}
\emailAdd{mauro@susy.house}

\abstract{The residential sector is responsible for approximately 20\% of the world's energy-related greenhouse gas emissions. In order to improve this reality, it is essential to help and support the leading actor in this scene, the homeowner. In general, buildings are complex objects as each home may have very particular parameters: different materials (windows, walls, types of roofs, etc.), uses, and habits. No two cases are identical. In this paper, we addressed this problem by proposing a method to obtain estimates of energy and carbon savings resulting from a class of retrofit projects and housing profiles. We applied our formulas to conduct a case study of the London residential sector, supported by the publicly available Energy Performance Certificates (EPC) dataset. As a result of a multi-project renovation of a house, including loft insulation, double-glazed windows, and energy-efficient lighting (LED), we found that the average family in the Greater London area could save approximately 9 MWh in energy consumption, 1,700 kgCO2 in carbon emissions and more than \pounds 800 on the energy bills, over a year,  in contrast to a cost of around \pounds 9,000 (or a return on investment of 11 years). In another example, applying the method to calculate what is needed to install heat pumps in each house in London, a total of \pounds 16.9 billion would be spent, leading to savings of \pounds 450 million per year (return on investment of 37 years). Finally, an overall upgrade to loft insulation would cost London households \pounds 1.4 billion, generating \pounds 290 million in annual savings (a 5-year return on investment).
}


\begin{document} 
\maketitle
\flushbottom

\newpage
\section{Introduction}
\label{sec1}

Worldwide, Industry and Transport are responsible for 36\% and 23\%  of global $CO_2$ emissions, respectively \cite{Allwood}. Policies all around the world are trying to address these sectors by following a \textit{top-down} approach, i.e. by imposing new rules, sanctions and targets, such as the establishment of clean air zones or the ban of fossil fuel powered cars. On the other hand, the Residential sector is responsible for 17-21\% of energy-related greenhouse gas emissions \cite{bbc}. Similar policies, however, are less likely to be efficient in this last case. 

There are approximately 2.3 billion dwellings in the world, for a population of 7.9 billion \cite{arch_design}. Each household corresponds to different fabrics (windows, walls, roof materials, etc.) and different uses and habits (number of occupants, temperature per room, water use, etc.). There are no two identical cases. In addition, behaviors are subject to continuous change. In order to deal with this level of complexity, we may need a different twofold approach: (i) each dwelling must be known in its very details (its infrastructure and its daily energy demand), and (ii) the householder must be the decision-maker for any property improvements/retrofits. This combination shows that, for the Residential sector, tackling energy-related $CO_2$ emissions is a \textit{bottom-up} process, where end-users have to be involved and engaged to form a complex network of billions of moves towards a better and sustainable system. Once again, 2.3 billion families are responsible for almost a fifth of global $CO_2$ emissions and must act as decision makers to change this reality.

It turns out that these agents may not have the relevant information to make key decisions to improve their homes. Unlike the Industry or Services sectors, where business analysts are dedicated to collecting and managing data and applying solutions to protect and improve their company assets, this is unlikely to be the case for families. In other words, at the domestic level, individuals may be unable to analyze their homes in terms of energy efficiency and performance in sufficient detail, and assess what type of retrofit projects are best suited to their properties, with a reasonable return on investment. Therefore, a natural question that arises is how can we provide information and analytical tools to help owners with these “green project” investment decisions?

In this article, we address the above question rephrased as: How to obtain estimates of energy and carbon savings resulting from a class of retrofit projects and for a variety of housing profiles? In Section \ref{sec2} we show that these quantities can be given as a function of the energy demand, $E_0$ in the initial state, that is, prior to any update within the mentioned class. We approximate $E_0$ by the Ordinary Least Squares regression of energy consumption on house volume and controlled by home type, built form and age. Next, in Section \ref{secres}, we conducted a case study of the London residential sector, which comprises as many as 3.49 million dwellings (houses, flats, bungalows and park homes) \cite{london_gov}. We apply our expressions to a London typical house in order to obtain intervals - i.e. error propagation - of cost and savings in energy and carbon emissions. Furthermore, we use the central values present in the Energy Performance Certificates (EPC) dataset \cite{EPC} to obtain dispersion across the London area. We present our results as sample mean by London boroughs. The Section \ref{disc} is devoted to a brief discussion of our results. 

\section{Methodology}
\label{sec2}

In the scenario where the internal space of a house is kept at a constant temperature, the Roof, Exterior Walls, Windows  and Floor will work as independent  energy consuming elements, similar to lightbulbs of different power.  
It would be useful, therefore, if we could estimate the amount of energy that each of these elements require, which in this case shall depend on their physical properties (e.g. material thermal transmittance) and the temperature variation with external space. On what follows, we define a \textit{Bare Home} (BH) a particular dwelling non-insulated, with single-glazed windows, no LED\footnote{Light-emitting diode.} lighting and standard main heating system (no Heat Pumps). Evidently, characteristics such as floor area, height, age, type (House, Flat, etc), built form (detached, semi-detached, etc..) are allowed to vary in our analysis, while the \textit{Bare Home} only denotes the initial state of a dwelling, prior to any upgrade devoted to efficiency. We will see in the next sections that, for instance, a wall is classified as `non-insulated' if its thermal transmittance is above a certain threshold. The advantage of the \textit{Bare Home} concept relies on the possibility that, by using the large EPC dataset, we can obtain a fair estimate for the amount of energy required by each of these independent house components, or
‘infrastructural appliances’. As mentioned before, the BH refers to the initial state of a particular dwelling, i.e. its state at a time $t = t_0$. Therefore, we refer as the $t_0$ of a dwelling to its \textit{bare} stage, and to $t > t_0$ to its upgraded state, the upgrade belonging to any of the projects we will explore along this paper. 

Let us denote the total amount of energy required by a given BH as $E_0$. From \cite{Palmer} we find that incandescent Lighting must require $\alpha_L$ portion of $E_0$, where $\alpha_L \approx 0.03$. The energy demand for Lighting is given exclusively by the number of light bulbs, times their power and time interval on. Evidently, this amount of energy is fixed, independent if the house contain Loft Insulation or not. We extrapolate this method to estimate the energy consumption from different components of a residence. Suppose, for instance, we know that single-glazed Windows require $\alpha_W E_0$ to keep the BH temperature constant, while a roof must require $\alpha_I E_0$. Once more, these amounts are related with the heat loss through these surfaces, which should depend only on their physical properties and the difference of temperature between their interior and exterior sides, common to the real dwelling and the BH. Hence, single-glazed will require $\alpha_W E_0$ in spite of any other retrofit projects. In order to estimate savings with the help of $E_0$, it is important that we keep the it fixed. In other words, if the energy demand in a given time \textit{t} is $E$, we should not assume that $\alpha_W E$ is the amount of energy lost through single-glazed in this house, since $\alpha_W$ is a result of a BH analysis as a reference, while real building may already present upgrades that result in $E < E_0$.

The use of $E_0$ is convenient when specific details about the building infrastructure are missing or difficult to obtain. We will notice, for instance, that the precise quantity saved from Insulation will depend on total insulated area and temperature variation between internal and external space. Similarly, the Lighting savings expression depends on number of outlets, power and time interval of lights on. These are variables normally difficult to obtain, except if provided directly by the user. Nevertheless, by rewriting the expressions in terms of a single quantity, we overcome the information sensitivity with a relatively small loss in precision, depending on the quality and size of the particular sample we apply our statistical methods on.

We will see that $E_0$ must be re-scaled whenever the energy cost of a particular structure is not independent of the presence of upgrades in other channels. This is the case for heat pumps\footnote{A heat pump is a heat transfer device, that is, it does not generate heat, but uses a pump to conduct thermal energy through a refrigeration cycle.} (HP). The HP reduces the amount of energy required for heating, which will be different for a house with or without insulation. Thus, in order to estimate HP savings we will need to derive the portion of the real demand that is used for space heating.

In summary, we consider a set of energy drain elements, namely Windows, Roof, Wall, Floor and Lighting, of a residence at constant temperature. We refer to these elements belonging to the same class in the sense that a change in walls should not impact the energy demand by the roof in a constant temperature regime.  We will see that energy sa\-vings from upgrades in the above elements can be computed if we know their impact in the energy demand of a reference \textit{Bare Home}, which we explore in the next session. We then derive an estimate for the BH total demand $E_0$ by performing an Ordinary Least Square (OLS) regression of consumption over volume, home type, built form and age, from the EPC dataset. We notice that a Heat Pump analysis cannot be performed in the similar fashion, since its function is to compensate the heat loss given the house infrastructure, hence directly correlated to the heat transmittance of the walls, roof, floor, among other features.

\subsection{Energy Consumption and Savings}
We selected four house retrofit projects for our environmental and economic analysis. The multi-glazed windows and LED lighting upgrades are proper for all types of dwellings (Houses, Flats, Bungalows, Maisonettes and Park homes), while Loft Insulation and Heat Pumps are exclusive for Houses. Following our previous discussion, we will provide estimates for energy and carbon savings as a function of the total energy consumption $E_0$ at the initial state, $t = t_0$. The time $t_0$ indicates the period prior to any infrastructural upgrade that is within our set of projects, while $t > t_0$ indicates the interval after the installation of the particular project described in the subsection. The energy savings resulting from the investment to upgrade the efficiency of a dwelling element $P$ (where $P = $ Windows, Roof, Loft, Floor, Lighting, Heating, etc..) will then be given by the difference between the energy consumption from $P$ before and after the upgrade, i.e.
\begin{equation}\label{light}
E^P_S = E^P(t_0) - E^P(t > t_0)
\end{equation} 
where $E_S$ denotes Energy \textit{Savings}. For example, on average around $20\%$ of annual demand dedicated to the heating system will compensate the heat loss due to single glazed windows (\cite{Palmer}, from average heat loss per dwelling in 1970), while $60\%$ of total annual $E_0$ is devoted to heating, thus resulting in $E^W(t_0) \approx 0.12 E_0$. We must emphasize that the notation $E(t_0)$, i.e. energy as a function of $t_0$, is invoked merely to denote the dependence of the outcome on the variables existent at a specific time interval. In this sense, the time variable may as well be replaced by an index. 

\subsubsection{Insulation}
\label{sec4}

Without loss of generality, we present our analysis of Insulation applied to the Loft area. The estimates can be obtained given the heat flux ($\phi$) relation with the temperature variation ($\Delta T$) and roof thickness ($L$), i.e.
\begin{equation}\label{ins}
\phi = \kappa \frac{\Delta T}{L} 
\end{equation}  
Hence, the thermal conductivity $\kappa$ measures the slope of the flux dependence on the ratio between temperature variation (interior and exterior space) and the material length, in the flux direction\footnote{Note that, since $\phi$ has units of $energy\cdot(time\cdot area)^{-1}$, $\kappa$ must have units of $energy\cdot(time\cdot length\cdot temperature)^{-1}$.}. We can rewrite the above equation in terms of heat transfer, in kWh per year, as
\begin{equation}\label{ins2}
E^I(t_0) =  8.76 \frac{\kappa_r}{L_{r}} \cdot \Delta T \cdot A 
\end{equation}
where the sub-index $r$ refers to `roof', while the upper-index \textit{I} denotes the Insulation project. $\Delta T$ is measured in Kelvin ($K$), loft area A in sqm ($m^2$), thickness $L_{r}$ in meters ($m$). The time $t_0$ indicates the initial level of insulation, defined by the roof material whose thermal conductivity is given by $\kappa_r$ in $W/(m \cdot K)$. The $8.76$ factor converts the result into $kWh/year$. The $E^I(t_0)$ is the energy that needs to be compensated by the heating system in order to maintain the house temperature, or $\Delta T$, constant. In this sense, a possible estimate for the $kWh$ saved via house insulation can be given by considering the difference between the $\kappa$ indexes of two different materials, present before and after insulation. 

The addition of a layer of new material under the roof will correspond to an effective constant $\kappa(L_{i})$, i.e. a function of the added insulation provided by that layer of thickness $L_{i}$. At $t > t_0$, or after insulation, we have:
\begin{equation}\label{ins3}
E^I(t) =  8.76 \frac{\kappa(L_{i})}{L_{r} + L_{i}} \cdot \Delta T \cdot A 
\end{equation}
where $\kappa(L_{i})$ should depend on both roof and insulation material, such that $\kappa(L_{i} = 0) = \kappa_r$. On what follows we approximate the above expression for the exchanged energy by:
\begin{equation}\label{ins4}
E^I(t) =  8.76 \frac{\kappa_i}{(\kappa_i/\kappa_r)L_{r} + L_{i}} \cdot \Delta T \cdot A 
\end{equation}
where $\kappa_i$ denotes the thermal conductivity of the insulating material (i.e. the insulation layer added). Notice that, for $\kappa_r \gg \kappa_i$, the roof's presence becomes negligible compared with the insulation.
In summary, the energy in kWh, $E_S$, saved in the interval of one year is given by the difference:
\begin{equation}\label{ins5}
E^I_S \equiv E^I(t_0) - E^I(t) =  8.76 \left[\frac{\kappa_r}{L_{r}} - \frac{\kappa_i}{(\kappa_i/\kappa_r)L_{r} + L_{i}}\right] \cdot \Delta T \cdot A 
\end{equation}

\begin{figure}[tbp]
	\centering 
	\includegraphics[width=.85\textwidth]{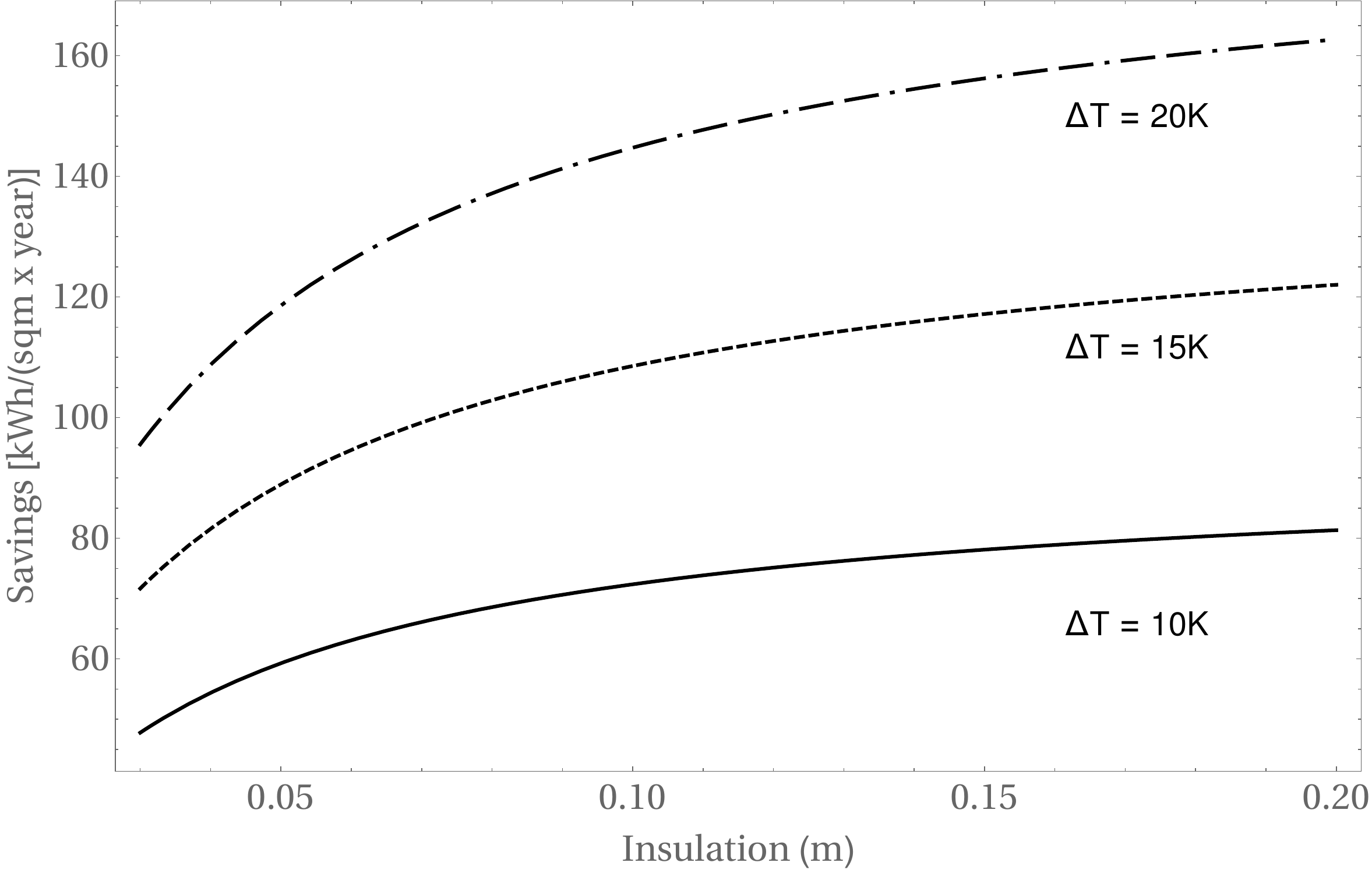}
	\caption{\label{fig:ins} Annual energy savings from Insulation, in kWh/sqm.}
\end{figure}

From \cite{Yang}, a typical roof without insulation would have a heat transfer coefficient of $\frac{\kappa_r}{L_r} = 1.06 W/(m^2K)$, while a standard insulation material will have $\kappa_i \approx 0.03 W/(m\cdot K)$. The Fig.\ref{fig:ins} shows annual energy savings per sqm, for three different values of $\Delta T$.  

Let us normalize the $E^I_S$ by $E^I(t_0)$, i.e.
\begin{equation}\label{ins6}
\frac{E^I_S}{E^I(t_0)} = \frac{\kappa_r L_{i}}{\kappa_i L_{r} + \kappa_r L_{i}} \end{equation}
Now, suppose that the fraction of the house total energy consumption, $E_0$, required for heating at $t = t_0$ previous to the insulation, is known and given by $E^I(t_0)= \alpha_I E_0$, where $0<\alpha_I<1$.  It follows that we can estimate $E^I_S$ as
\begin{equation}\label{ins7}
E^I_S = \alpha_I \frac{\kappa_r L_{i}}{(\kappa_i L_{r} + \kappa_r L_{i})}E_0 
\end{equation}
Notice that, if the loft area $A$ is known, $E_S^I$ is given directly from Eq.(\ref{ins5}). Alternatively, one can use Eq.(\ref{ins7}) where $\alpha_I \approx 0.06$ is the average heat loss by energy ratio per dwelling, prior to 1970, estimated by \cite{Palmer}.

\subsubsection{Windows}
\label{sec5}
The principles behind the outcome of a Windows project are analogous to those of the Insulation where now we consider the single $(I)$, double $(II)$ or triple $(III)$ glazed windows thermal transmittance. The annual energy required to maintain constant the difference of temperature $\Delta T$ between external and internal environment due to the heat exchange through the windows of area $A$ can be given by:
\begin{equation}\label{wind}
E^W(t) = 8.76 \cdot U_{(I,II,III)} \cdot \Delta T \cdot A 
\end{equation}
where $U$ is the thermal transmittance, in $W/(m^2\cdot K)$. The U-value of single and multi-glazed windows is given by  $U_I = 5.74$, $U_{II} = 2.7$ and $U_{III} = 0.7$ \cite{Jorge}. Suppose that $(1- \lambda)$ is the proportion of windows area to be replaced, from single to double glazed. The annual amount of energy saved can be estimated by:
\begin{equation}\label{wind2}
E^W_S = 8.76 (1- \lambda)(U_I - U_{II}) \cdot \Delta T \cdot A
\end{equation}
As we mentioned before, the amount of energy required by single-glazed windows is approximately $12\%$ of the total demand in a \textit{Bare Home} \cite{Palmer}. It follows that, at $t = t_0$, 
\begin{equation}\label{wind3}
E^W(t_0) = (1- \lambda)\cdot\alpha_W\cdot E_0
\end{equation}
where $\alpha_W \approx 0.12$. By normalizing $E^W_S$ by $E^W(t_0)$ and considering the above expression, we can rewrite the energy savings as
\begin{equation}\label{wind4}
E^W_S = (1- \lambda)\cdot\alpha_W\cdot E_0 \cdot \frac{(U_I - U_{II})}{U_I}
\end{equation}
Notice that one can also obtain an estimate of the total single-glazed windows area from Eq.(\ref{wind}) and Eq.(\ref{wind3}):
\begin{equation}\label{wind5}
A \approx \frac{1 - \lambda}{8.76} \frac{\alpha_W E_0}{U_I \Delta T} 
\end{equation}
The expression above will be useful to help us estimate the total cost of a Windows retrofit. We must emphasize that $\lambda$ is one of the variables present at the EPC dataset.

\subsubsection{Lighting}
\label{sec6}
In this section we develop an estimate for the amount of energy saved by replacing high energy and low lumen (incandescent) light bulbs to low energy high lumen LED lamps. 
First, suppose all the lighting in the dwelling is generated by $N$ inefficient light bulbs, whose corresponding energy consumption is a fraction $\alpha_L$ of the the total BH reference $E_0$, i.e.
\begin{equation}\label{light2}
E^L(t_0) = \alpha_L E_0 = N \cdot P \cdot \Delta t
\end{equation}
where $P$ is the power of the light bulbs and $\Delta t$ is the time interval of lights on. Next, consider the more general case where, at $t = t_0$, there are $m$ low-energy light bulbs already installed in the house, whose power is a fraction $\beta$ of $P$. In this scenario,
\begin{equation}\label{light3}
E^L(t_0) = \left(m \beta P + (N-m)P\right)\Delta t 
\end{equation} 
or 
\begin{equation}\label{light4}
E^L(t_0) = \left(1 - (1-\beta) \frac{m}{N}\right) N P \Delta t 
\end{equation} 
After replacing all the inefficient lamps for LED bulbs, the energy consumption will be given by:
\begin{equation}\label{light5}
E^L(t > t_0) = \beta N P \Delta t 
\end{equation}
By replacing Eq.(\ref{light4}) and Eq.(\ref{light5}) appropriately in Eq.(\ref{light}), we estimate the Lighting energy savings by:
\begin{equation}\label{light6}
E^L_S = (1-\beta)\left(1 - \frac{m}{N}\right) N P \Delta t 
\end{equation}
Finally, from Eq.(\ref{light2})
\begin{equation}\label{light7}
E^L_S = \alpha_L(1-\beta)\left(1 - \frac{m}{N}\right) E_0 
\end{equation}
In the next section, we take $\alpha_L \approx 0.03$ \cite{Palmer} and $\beta = 0.25$ \cite{Alshorman}. The proportion of low energy lighting bulbs in a building is denoted by the ratio $m/N$, which has been measured and recorded as a variable in the EPC table. 

\subsubsection{Heat Pumps}
\label{sec7}
Heat pumps (HP) are an energy efficient and low-carbon option for space heating. Their corresponding energy savings ($E_S$) are estimated to be around 60-75\% of the demand from gas boilers \cite{gov}. As in Eq.(\ref{light}), we can formally write $E^H_S$ as the difference in consumption before and after the implementation of a HP project. 
From \cite{Palmer}, $E^H(t_0) \approx  0.6E_0$, i.e. the annual energy cost of heating is on average near 60\% of the total energy requested by the householders. We can then estimate $E^H_S$ from HPs as
\begin{equation}\label{HP}
E^H_S = (1 - \lambda)\alpha^H E_0
\end{equation}
where we use $\lambda = 0.25$ and $\alpha^H \approx 0.6$. Notice that the HP savings are a fraction of the space heating energy. Hence, in a multi-project setup including heating related upgrades, such as multi-glazed Windows and Insulation, we must replace $\alpha^H E_0 \rightarrow \alpha^H E_0 - E^W_S - E^I_S$. In other words, HP savings are posterior to the savings from actual Windows and Insulation profile.

Since the source of energy for heating at $t_0$ can be, mainly, either Gas or Electricity, one cannot consider a single conversion factor to compute our Carbon and GBP\footnote{Great Britain Pound or British pound sterling, the official currency of the United Kingdom, symbolized by the pound sign \pounds.} savings estimates. From Eq.(\ref{HP}) we have:
\begin{equation}\label{HP2}
CO2_S = (\gamma_C(t_0) - \lambda \gamma_{C;EL})\alpha^H E_0
\end{equation}
where $CO2_S$ is the savings in Carbon emissions and $\gamma_{C;EL}$ denotes the conversion factor between Electricity consumption and $CO2$. At $t_0$ we may have $\gamma_C(t_0) = \gamma_{C;EL}$ or $\gamma_C(t_0) = \gamma_{C;GAS}$, depending on the energy source. Analogously, savings in GBP (\pounds) are derived from a conversion factor $\gamma_\pounds$, i.e.:
\begin{equation}\label{HP3}
\pounds_S = (\gamma_\pounds(t_0) - \lambda \gamma_{\pounds;EL})\alpha^H E_0
\end{equation} 
Notice, therefore, that GBP savings are positive when $\lambda \gamma_{\pounds;EL} < \gamma_\pounds(t_0)$. At the present moment\footnote{\today.} we observe $\gamma_{\pounds;EL} = 0.34 (\pounds/kWh)$, such that $\lambda \gamma_{\pounds;EL} \approx 0.08 (\pounds/kWh)$. In the case of natural Gas we find  $\gamma_{\pounds;Gas} = 0.1 (\pounds/kWh)$. Note, however, that these values represent a generic estimation rather than a precise expression of their market values.

\subsection{\texorpdfstring{$E_0$}{TEXT} OLS Regression}\label{e0reg}

We adopt a simple linear regression model where the dwelling \textit{volume} is taken as the single explanatory variable to the energy consumption outcome, having \textit{home type}, \textit{built form}, and \textit{age} as controlling variables. The model is inspired by the correspondence between the temperature of an ideal gas and the volume at constant pressure, as well as the linear dependence of thermal energy and temperature, assuming that the dwelling ``effective specific heat capacity'' is constant in all controls. Alternative models have been discussed in Section \ref{disc}. Here, we write our regression formula in the standard form:
\begin{equation}\label{OLS}
\overline{E_0} = \beta_0 + \beta_1 V + \sum_j \lambda_j C_j + \xi 
\end{equation}
The `bar' on the outcome variable reminds us that the observations may be `polluted' by dwellings not in the \textit{Bare Home} class. Our final estimate will be a result of re-scaling $\overline{E_0}$ with infrastructural variables reported in the EPC. $\beta_0$ is the intercept, while $\beta_1$ is the slope of the dependence on volume, $V$. The $\lambda_j$ are coefficients of the $j$ control variables, $C_j$. The `error term' $\xi$ takes into account potentially relevant variables neglected in our model. The OLS estimates of the model coefficients are presented in Table \ref{tab1}. We have excluded the categories\footnote{The domain of the \textit{Home type} variable is given by [House, Flat, Bungalow, Maisonette, Park Home], while for \textit{Built Form} we have [Detached, Semi-Detached, End-Terrace, Mid-Terrace, Enclosed End-Terrace, Enclosed Mid-Terrace].} \textit{House}, \textit{Detached}, and buildings built prior to 1900, in order to prevent multicollinearity. Our data sample comprises 2,191,399 London dwellings. In order to obtain the \textit{Volume} variable, we considered the average floor height from values within the interval of $1m$ to $10m$. In addition, the floor area was constrained to be greater than $10m^2$. Results of average floor height, in meters, are\footnote{The notation $y(\pm x)$ indicates the interval created by $x$ and the final significant digits of $y$. For example, $1.25(\pm0.15) \equiv [1.1, 1.4]$.}: $2.53(\pm0.22)$ (House), $2.53(\pm0.26)$ (Flat), $2.52(\pm0.27)$ (Maisonette), $2.49(\pm0.19)$ (Bungalow) and $2.55(\pm0.24)$ (Park home). We consider $\overline{E_0}$ in units of kWh per month (annual EPC variable over 12 months). We compute the Mean Squared Error (MSE) from a 10-fold cross-validation, resulting in 
$\sqrt{MSE} = 675(\pm170)$ $kWh/month$ from table \ref{tab1}.   

As we mentioned before, the model of Eq.(\ref{OLS}) is inspired by the linear dependence of temperature on volume for an ideal gas. We can also test the accuracy of a model representing the van der Waals equation for non-ideal gases, where $T \approx \beta_0 + \beta_1 V + \beta_2 V^{-1} + \beta_3 V^{-2} $. In this scenario, we have obtained $\sqrt{MSE} = 674(\pm155)$ $kWh/month$. For simplicity, we adopt the results derived from Eq.(\ref{OLS}).

$E_0$ estimation, in principle, should rely exclusively on data of dwellings whose infrastructure are in accordance with our \textit{Bare Home} definition. In principle we could select a BH sample from our original dataset. However, the absence of a project (i.e. multi-glazed windows, LED, etc..) in a building might be correlated with specific features that limited the project installation in the first place. The correlation may be impossible to be measured if these features were not taken into account in the dataset. For example, a heat pump may not be viable for a building due to limitations in the external area. Similarly, the presence of multi-glazed windows may depend on hidden economical variables. Since we consider the observations in the full dataset, $\overline{E_0}$ is affected by the actual presence of the selected upgrades in the London buildings. In order to obtain an $E_0$ estimate, we re-scale our result by
\begin{equation}\label{E0}
\overline{E_0} \rightarrow E_0 = \overline{E_0}\left(1 - \alpha_W \lambda\frac{U_I - U_{II}}{U_I} - \alpha_L (1-\beta)\frac{K}{N} - \alpha_I \frac{\kappa_r L_{i}}{(\kappa_i L_{r} + \kappa_r L_{i})}\right)^{-1}
\end{equation}
We estimate the $\lambda$, $K/N$ and $L_i$ parameters as the mean presence of multi-glazed windows, lighting and insulation (cm), respectively, over the observations. A House in London will have, on average, $9.2(\pm9.6)$ cm of Loft Insulation, $85(\pm30)\%$ of its Windows multi-glazed, and 53(36)\% of its lightbulbs of LED type. A building of different type will have 78(39)\% of its Windows multi-glazed and 60(38)\% of its lightbulbs of LED type. We have selected observations whose main heating system is other than heat pumps. In fact, only 0.15\% of London houses use HPs.

\begin{table}[t]\centering
	{\footnotesize\begin{tabular}{lrrrrrr}
			\toprule
			{} &    Coef. &  Std.Err. &        t &  P>|t| &   [0.025 &   0.975] \\
			\midrule
			$\beta_0$        &  1048.27 &      3.14 &   334.04 &   0.00 &  1042.12 &  1054.42 \\
			Volume  &     6.36 &      0.01 &  1235.61 &   0.00 &     6.35 &     6.37 \\
			Flat  &  -232.51 &      1.41 &  -165.23 &   0.00 &  -235.27 &  -229.75 \\
			Bungalow  &    91.52 &      4.69 &    19.53 &   0.00 &    82.34 &   100.71 \\
			Maisonette  &  -209.86 &      2.37 &   -88.42 &   0.00 &  -214.51 &  -205.21 \\
			Park Home  &   245.74 &    180.50 &     1.36 &   0.17 &  -108.03 &   599.52 \\
			Semi-Detached &  -159.43 &      2.15 &   -74.00 &   0.00 &  -163.65 &  -155.21 \\
			End-Terrace &  -205.89 &      2.22 &   -92.83 &   0.00 &  -210.23 &  -201.54 \\
			Mid-Terrace &  -403.92 &      2.00 &  -201.49 &   0.00 &  -407.85 &  -399.99 \\
			Enclosed End-Terrace &  -218.63 &      3.84 &   -56.86 &   0.00 &  -226.16 &  -211.09 \\
			Enclosed Mid-Terrace &  -318.19 &      3.88 &   -81.96 &   0.00 &  -325.80 &  -310.58 \\
			1900-1929   &   -70.09 &      1.93 &   -36.39 &   0.00 &   -73.86 &   -66.31 \\
			1930-1949   &  -161.57 &      2.02 &   -80.11 &   0.00 &  -165.52 &  -157.62 \\
			1950-1966   &  -304.83 &      2.26 &  -135.09 &   0.00 &  -309.25 &  -300.41 \\
			1967-1975   &  -316.94 &      2.47 &  -128.56 &   0.00 &  -321.77 &  -312.11 \\
			1976-1982   &  -482.46 &      3.15 &  -153.05 &   0.00 &  -488.64 &  -476.28 \\
			1983-1990   &  -460.52 &      3.02 &  -152.65 &   0.00 &  -466.43 &  -454.60 \\
			1991-1995   &  -486.21 &      3.60 &  -135.24 &   0.00 &  -493.26 &  -479.17 \\
			1996-2002   &  -674.63 &      3.14 &  -214.79 &   0.00 &  -680.78 &  -668.47 \\
			2003-2006  &  -814.28 &      3.31 &  -245.74 &   0.00 &  -820.78 &  -807.79 \\
			2007-2011  &  -917.81 &      3.50 &  -262.13 &   0.00 &  -924.67 &  -910.94 \\
			2012-2022  & -1415.19 &      4.37 &  -324.05 &   0.00 & -1423.75 & -1406.63 \\
			\bottomrule
	\end{tabular}}\caption{Regression of Energy Consumption on building Volume.}\label{tab1}
\end{table}

In summary, we regress Energy consumption on volume, controlling over home type, built form and age in order to estimate $\overline{E_0}$. Finally, we find average parameters representing the actual presence of efficiency elements, and re-scale our reference energy by the transformation of Eq.(\ref{E0}). Results correspond to an increase of $11\%$, (i.e. $\overline{E_0} \rightarrow E_0 = 1.11\overline{E_0}$) in the estimate of $E_0$ for Houses, and $5\%$ increase in $E_0$ for the remaining home types.

\subsection{Carbon Footprint and Savings}

The amount of savings in carbon dioxide emissions is obtained with the aid of the \textit{Global Warming Potential} (GWP) concept \cite{GWP}, which considers the $CO_2$ potential for absorbing thermal radiation  as a reference for different types of gas. For example, 1 kWh generated from natural gas is considered to have an impact on global warming equivalent to 0.184 kg of $CO_2$ released into the atmosphere \cite{Carbon}. In particular, the GWP for grid\footnote{Grid: A network of conductors for distribution of electrical power (Merriam-Webster).} electricity will depend on the different means of generation adopted by the specific country. We consider that 1 kWh of energy provided by the grid will be equivalent to 0.20kg of $CO_2$ emissions \cite{Carbon}. The use of these conversion factors must also depend on the infrastructural element that the retrofit addresses. In a Lighting project, we must naturally convert $E_S^L$ into $CO_2$ equivalent by invoking the Grid conversion factor. On the other hand, the Insulation upgrade is expected to impact the energy devoted to space heating, so that the $E_S^I \rightarrow kgCO_2$ relation must depend on the building main heating system.

\section{Results}
\label{secres}

We now apply the expressions derived in Section \ref{sec2} to a collection of different configurations in the London boroughs, obtained from the EPC data set. First, we select a reference building, to be defined in the next section, whose controlled profile corresponds to the most frequent occurrence in the EPC table. Our goal is to perform error propagation analysis given the deviation of cost and physical parameters commonly found in the literature. Next, we assume the central values of the basic parameters in our calculation to study the dispersion of savings and cost of the retrofit projects in the London buildings. We finally consider the ratio of houses to estimate absolute values of project cost and savings per London borough.

\begin{figure}[hbp]
	\centering 
	\includegraphics[width=0.8\linewidth]{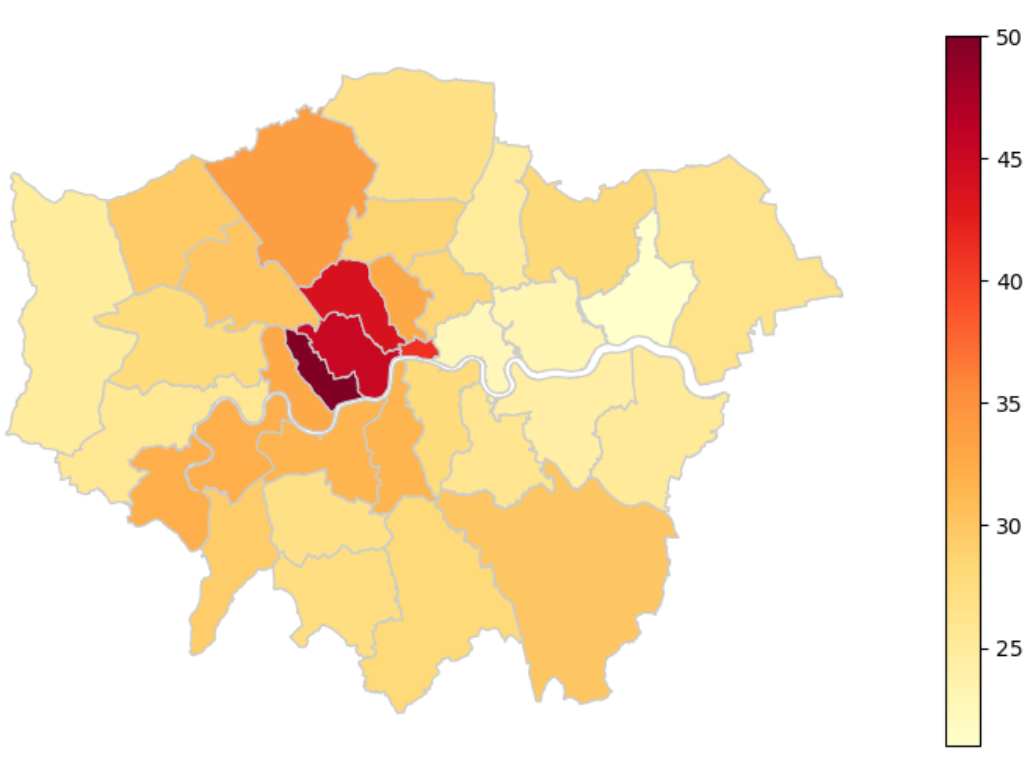}
	\caption{\label{fig:ii} Total annual energy consumption (MWh) of London Houses, per Borough.}
\end{figure}

\begin{figure}[hbp]
	\centering 
	\includegraphics[width=0.8\textwidth]{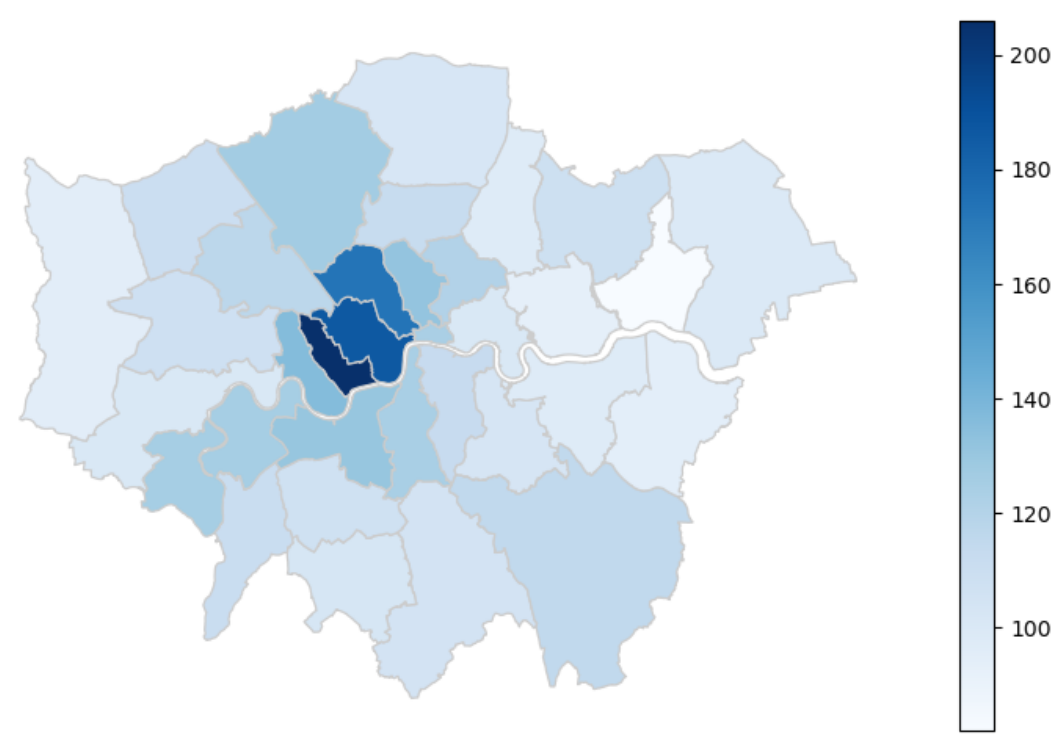}
	\caption{\label{fig:viii} Average floor area ($m^2$) of London Houses, per Borough.}
\end{figure}

\begin{figure}[hbp]
	\centering 
	\includegraphics[width=0.8\textwidth]{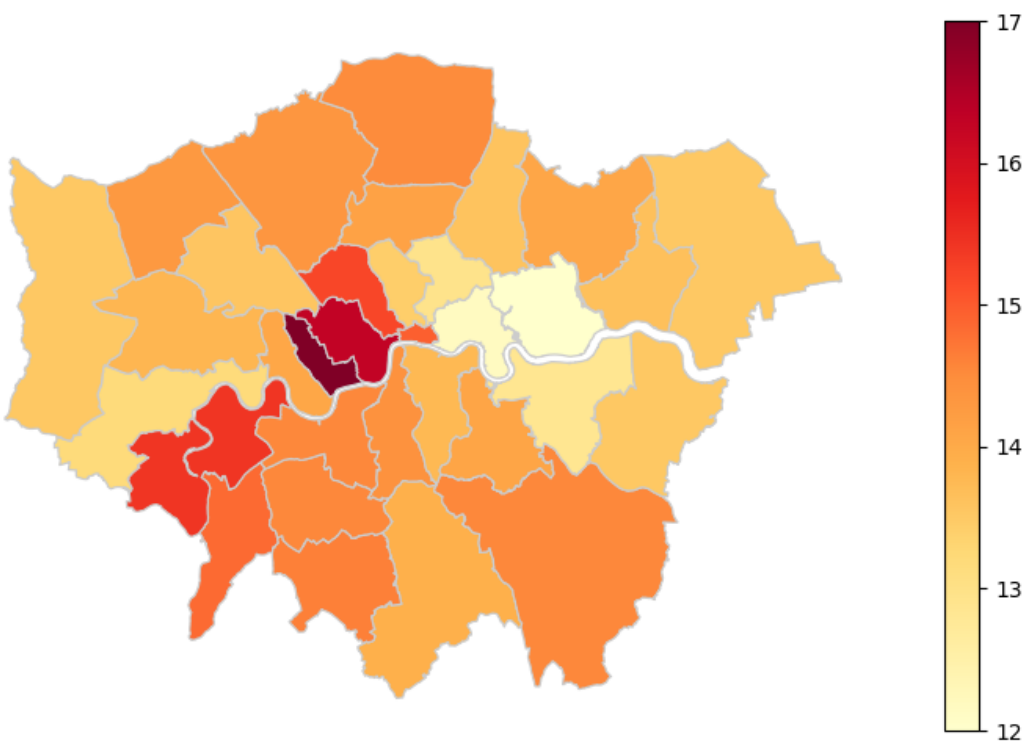}
	\caption{\label{fig:i} Total annual energy consumption (MWh) of London Flats, per Borough.}
\end{figure}

\begin{figure}[hbp]
	\centering 
	\includegraphics[width=0.8\textwidth]{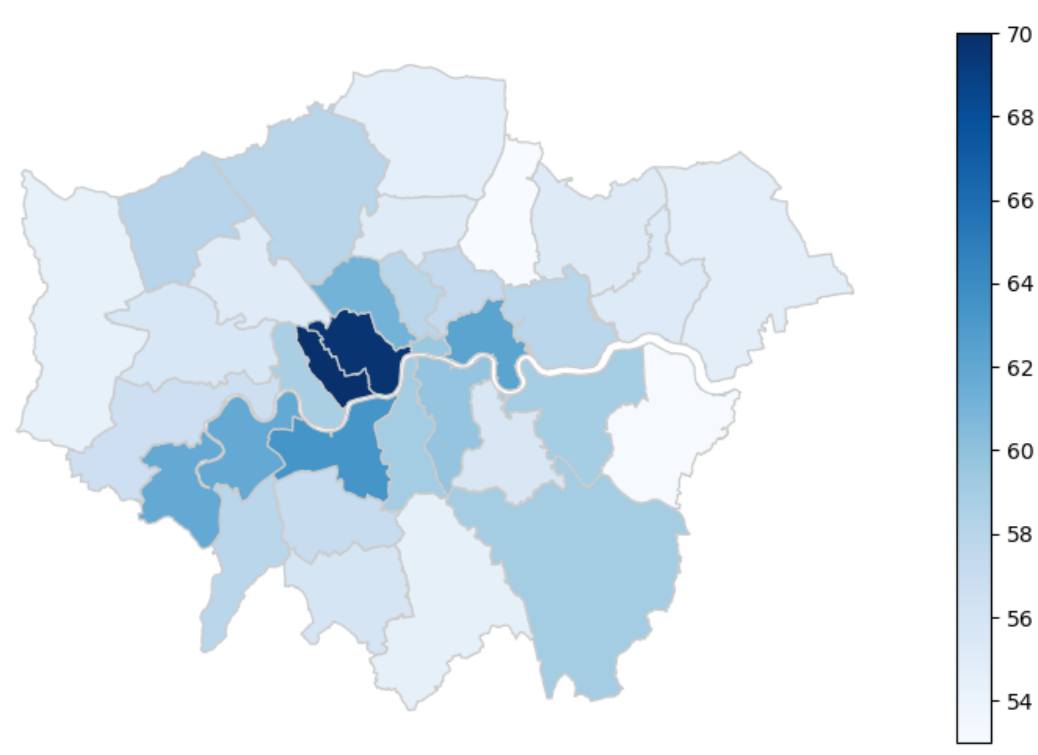}
	\caption{\label{fig:ix} Average floor area ($m^2$) of London Flats, per Borough.}
\end{figure}

Before we move to the report of our results, we must remember that Lighting, Windows, Loft Insulation (LI) and Heat Pumps (HP) correspond to the four retrofit strategies we  focused in our analysis. Notice that the first two apply to all types of dwellings, while the remaining ones are exclusive to Houses. In fact, according to EPC, 37\% of London residences are classified as houses, versus 55\% of flats. The Table 
\ref{house} illustrates a typical form that we can present some relevant statistics. The first column refers to the annual energy consumption, in kWh, of London houses, while the second presents the consumption per unit of floor area (sqm), and the last column the total floor area. The \textit{count} row shows the number of observations (810639) considered in the computation of the \textit{mean} and the standard deviation, \textit{std}. The remaining rows represent minimum, maximum and quartiles. Hence, the sample mean of annual consumption, given that the dwelling is a House, is $29(\pm17$) MWh. Average floor area is equivalent to $110(\pm60)$ sqm, while average consumption density of $262(\pm92)$ kWh/sqm. Evidently that in this analysis, each observation corresponds to a single building and, therefore, the large standard deviation reveals the dispersion in consumption among all residences, which must be impacted by features such as area, location and economical variables. In order to explore the dispersion, we will provide heat maps similar to Figure \ref{fig:ii}, where we compute the sample mean grouped by the London Borough. We see, for instance, how the demand per house in Kensington and Chelsea is approximately twice the value in Barking and Dagenham. In Figure \ref{fig:viii} we plot the equivalent map for the total floor area, indicating a strong correlation between consumption and space. The Table \ref{flat}, and Figures \ref{fig:i} and \ref{fig:ix}, show the Flat results, from a greater sample size of 1.2 million of observations. Despite lower demand, consumption per floor area of flats is compatible with that of houses. Similar to the previous case, maps indicate strong positive correlation between demand and space, with few exceptions, e.g. in Tower Hamlets and Lewishan. In general, the heat maps in the last subsection shall be interpreted  as the average of energy savings, per dwelling, resulting from the investment on given  project and at the particular borough. We must notice that a map of uniform texture indicates low dependence of savings on floor area.

\begin{table}[h]\centering
	{\footnotesize\begin{tabular}{lrrr}
			\toprule
			{} &  Annual Consumption (kWh) &  Density ($kWh/m^2$) &  Floor area ($m^2$) \\
			\midrule
			Sample Size &             810639 &                   810639 &      810639 \\
			Mean  &              28126.48 &                      261.67 &         109.84 \\
			Std   &              16959.79 &                       91.88 &          60.07 \\
			Min   &                 65.00 &                        1.00 &          10.08 \\
			25\%   &              18590.00 &                      203.00 &          79.72 \\
			50\%   &              24381.00 &                      248.00 &          96.00 \\
			75\%   &              32784.58 &                      302.00 &         123.00 \\
			Max   &            3298651.25 &                    11626.00 &       10815.25 \\
			\bottomrule
	\end{tabular}}
	\caption{London Houses: energy consumption per unit/year and per unit/sqm, considering a total of 810,639 observations. Above the sample mean, standard deviation and quartiles.}\label{house}
\end{table}

\begin{table}[h]\centering
	{\footnotesize\begin{tabular}{lrrr}
			\toprule
			{} &  Annual Consumption (kWh) &  Density ($kWh/m^2$) &  Floor area ($m^2$) \\
			\midrule
			Sample Size &            1203335 &                  1203335 &     1203335 \\
			Mean  &              14028.79 &                      255.13 &          59.04 \\
			Std   &               8067.55 &                      132.33 &          27.93 \\
			Min   &                 39.59 &                        1.00 &          10.01 \\
			25\%   &               9026.46 &                      172.00 &          45.00 \\
			50\%   &              12495.00 &                      231.00 &          55.50 \\
			75\%   &              17123.80 &                      310.00 &          69.10 \\
			Max   &             971152.20 &                    16464.00 &        5029.00 \\
			\bottomrule
	\end{tabular}}
	\caption{London Flats: energy consumption per unit/year and per unit/sqm, considering a total of 1,203,335 observations. Above the sample mean, standard deviation and quartiles.}\label{flat}
\end{table}

\begin{figure}[hbp]
	\centering 
	\includegraphics[width=0.9\linewidth]{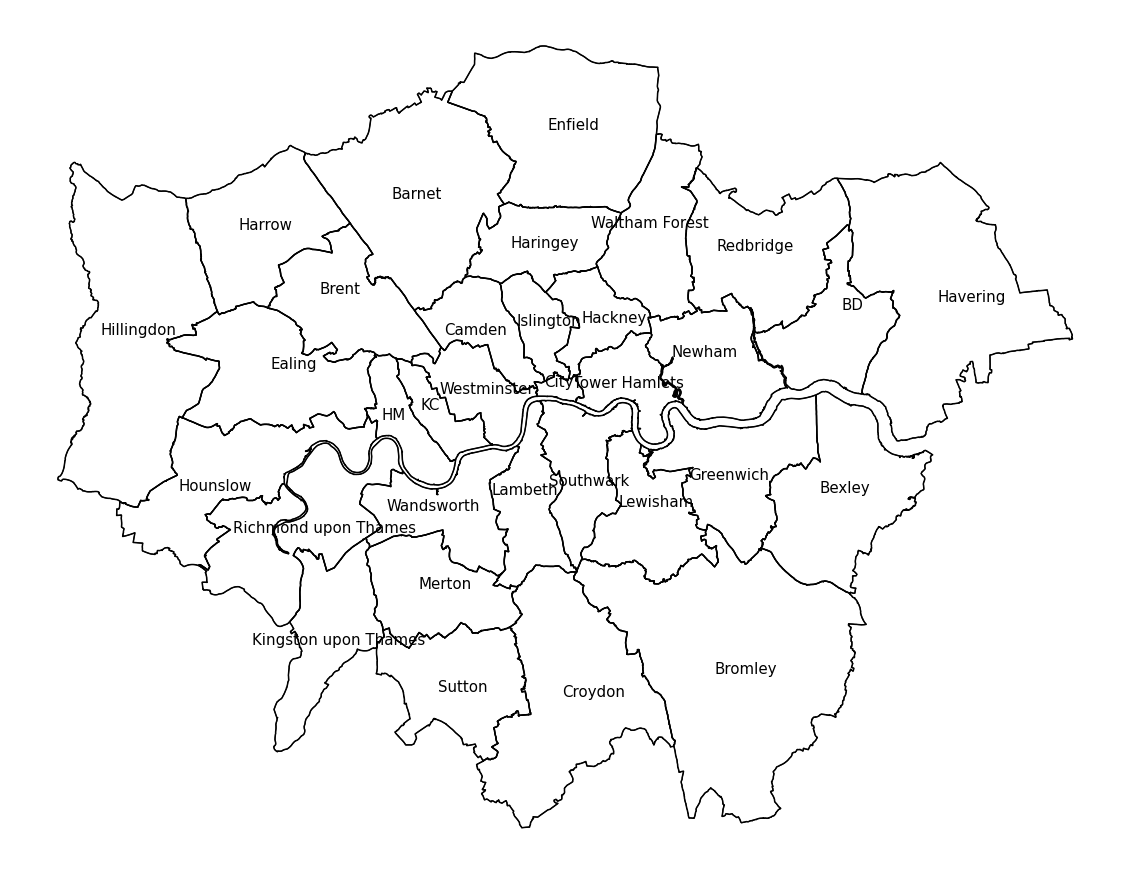}
	\caption{\label{fig:name} London Borough Labels. HM = Hammersmith and Fulham, KC = Kensington and Chelsea,
 BD = Barking and Dagenham, City = City of London.}
\end{figure}

\clearpage

\subsection{Sample House: Definitions, List of Parameters and Summary of Results}
\label{subsec1}

In this section we adopted the properties of a residence with the most frequent label of  built form and age among London houses. Here, a `\textit{Bare Home}' (BH) is a semi-detached house, built between 1930 and 1949, with 109sqm of floor area, and gas main heating system. To estimate a range for our outputs - i.e. their standard deviation (std) - we perform error propagation on the following variables: 

\begin{itemize}
	\item Annual Average of External Temperature: $12(\pm2)$ C
	\item Gas cost: $0.08(\pm0.01)$ GBP/kWh
	\item Electricity cost: $0.30(\pm0.01)$ GBP/kWh
	\item Loft Insulation Cost: $1.5(\pm0.5)$ GBP/(sqm*cm)
	\item Loft Insulation Cost (Installation): $15(\pm5)$ GBP/sqm
	\item Windows Cost (Installation): $120(\pm20)$ GBP/day
	\item Double-Glazed Windows Cost (Material): $500(\pm100)$ GBP/sqm
	\item Windows U-value (single-glazed): $5.7(\pm0.7)$ W/(sqm K)
	\item Windows U-value (double-glazed): $2.7(\pm0.7)$ W/(sqm K)
	\item Heat Pump (Cost + Installation): $11000(\pm2000)$ GBP
	\item LED light bulb cost: $7(\pm2)$ GBP
	\item Annual Energy Consumption: $29530(\pm28)$ kWh
	\item Annual $CO_2$ Emissions: $5906(\pm6)$ $KgCO_2$
\end{itemize}

The values in parentheses represent the standard deviation. For installation and material cost we assume an illustrative range of 10\% to 20\% of the central value. The remaining constants were collected from the literature. Bare House's annual energy consumption is the result of a linear regression of demand on volume, controlled by house type, construction form and age, of over two million homes in the London area. Annual Carbon Emissions is the result of conversion from Energy assuming 60\% of total demand in Gas, while 40\% in Electricity. In the next section, the \textit{BH} infrastructural profile consists of single-glazed windows, inefficient (incandescent) light bulbs and no presence of loft insulation, solar panels or heat pumps.

In Tables (\ref{insD})-(\ref{hpMD}) below we can find the description of four different retrofit project scenarios. When obtaining the outcome of a combination of these projects, the results are additive, except in the presence of heat pumps.  Our estimates of specific retrofit strategies for the selected BH can be summarized as follows:

\begin{description}
	\item[A] Double Glazed Windows + Loft Insulation (15cm):
	\begin{description}
		\item[Annual Savings (kWh):] $8439 \ \ (\pm 1693)$ 
		\item[Annual Savings ($KgCO_2$):] $1553 \ \ (\pm 311)$
		\item[Annual Savings (\pounds):] $614 \ \ (\pm 123)$ 
		\item[Total Cost (\pounds):] $8873 \ \ (\pm 2059)$ 
		\item[Return on Investment (years):] $ 14 \ \ (\pm 4)$  
		\item[Reduction in Energy Demand (\%):] $29 \ \ (\pm 6)$ 
		\item[Reduction in $CO_2$ Emissions (\%):] $26 \ \ (\pm 5)$
	\end{description}
	\item[B] Loft Insulation (15cm) + Double Glazed Windows + Low Energy Lighting:
	\begin{description}
		\item[Annual Savings (kWh):] $9104 \ \ (\pm 1693)$ 
		\item[Annual Savings ($KgCO_2$):] $1705 \ \ (\pm 311)$
		\item[Annual Savings (\pounds):] $813 \ \ (\pm 123)$ 
		\item[Total Cost (\pounds):] $9000 \ \ (\pm 2059)$ 
		\item[Return on Investment (years):] $11 \ \ (\pm 3)$
		\item[Reduction in Energy Demand (\%):] $30 \ \ (\pm 5)$ 
		\item[Reduction in $CO_2$ Emissions (\%):] $29 \ \ (\pm 5)$
	\end{description}
	\item[C] Loft Insulation (15cm) + Double Glazed Windows + Heat Pumps:
	\begin{description}
		\item[Annual Savings (kWh):] $15402 \ \ (\pm 2128)$ 
		\item[Annual Savings ($KgCO_2$):] $2690 \ \ (\pm 376)$
		\item[Annual Savings (\pounds):] $711 \ \ (\pm 140)$ 
		\item[Total Cost (\pounds):] $19897 \ \ (\pm 2873)$ 
		\item[Return on Investment (years):] $28 \ \ (\pm 7)$
		\item[Reduction in Energy Demand (\%):] $52 \ \ (\pm 7)$ 
		\item[Reduction in $CO_2$ Emissions (\%):] $46 \ \ (\pm 6)$
	\end{description}
	\item[D] Double Glazed Windows + Low Energy Lighting + Loft Insulation + Heat Pumps:
	\begin{description}
		\item[Annual Savings (kWh):] $16066 \ \ (\pm 2128)$ 
		\item[Annual Savings ($KgCO_2$):] $2843 \ \ (\pm 376)$
		\item[Annual Savings (\pounds):] $910 \ \ (\pm 141)$ 
		\item[Total Cost (\pounds):] $20025 \ \ (\pm 2874)$
		\item[Return on Investment (years):] $22 \ \ (\pm 5)$ 
		\item[Reduction in Energy Demand (\%):] $54 \ \ (\pm 7)$ 
		\item[Reduction in $CO_2$ Emissions (\%):] $48 \ \ (\pm 6)$
	\end{description}
\end{description}

\begin{table}[h]\centering
	{\footnotesize\begin{tabular}{lrrrr}
			\toprule
			{} &  Annual Savings (kWh) &  Annual Savings (KgCO2) &  Annual Savings (GBP) &  Cost (GBP) \\
			\midrule
			Sample Size &           1000 &             1000 &           1000 &   1000 \\
			Mean  &           6906.12 &             1270.72 &            502.56 &   4059.35 \\
			Std   &           1644.80 &              302.65 &            119.69 &    965.64 \\
			Min   &           1829.00 &              337.00 &            133.00 &    828.00 \\
			25\%   &           5785.00 &             1064.50 &            421.00 &   3383.00 \\
			50\%   &           6950.00 &             1278.50 &            506.00 &   4065.00 \\
			75\%   &           8005.25 &             1473.00 &            583.00 &   4748.00 \\
			Max   &          12098.00 &             2226.00 &            880.00 &   6768.00 \\
			\bottomrule
	\end{tabular}}
	\caption{Bare House 15cm Loft Insulation Upgrade Cost and Savings.}\label{insD}
\end{table}

\begin{table}[h]\centering
	{\footnotesize \begin{tabular}{lrrrr}
			\toprule
			{} &  Annual Savings (kWh) &  Annual Savings (KgCO2) &  Annual Savings (GBP) &  Cost (GBP) \\
			\midrule
			Sample Size &           1000 &             1000 &           1000 &   1000 \\
			Mean  &           1533.44 &              282.14 &            111.59 &   4813.17 \\
			Std   &            401.57 &               73.89 &             29.24 &   1818.26 \\
			Min   &             52.00 &               10.00 &              4.00 &   1800.00 \\
			25\%   &           1277.00 &              235.00 &             93.00 &   3599.75 \\
			50\%   &           1560.00 &              287.00 &            114.00 &   4495.50 \\
			75\%   &           1801.00 &              331.00 &            131.00 &   5511.75 \\
			Max   &           2871.00 &              528.00 &            209.00 &  16664.00 \\
			\bottomrule
	\end{tabular}}
	\caption{Bare House Multi-Glazed Windows Upgrade Cost and Savings.}\label{windD}
\end{table}

\begin{table}[h]\centering
	{\footnotesize\begin{tabular}{lrrrr}
			\toprule
			{} &  Annual Savings (kWh) &  Annual Savings (KgCO2) &  Annual Savings (GBP) &  Cost (GBP) \\
			\midrule
			Size &           1000 &             1000 &           1000 &   1000 \\
			Mean  &            664.44 &              152.93 &            199.20 &    127.60 \\
			Std   &              0.68 &                0.25 &              6.81 &     36.66 \\
			Min   &            662.00 &              152.00 &            174.00 &     24.00 \\
			25\%   &            664.00 &              153.00 &            195.00 &    103.75 \\
			50\%   &            664.00 &              153.00 &            199.00 &    127.00 \\
			75\%   &            665.00 &              153.00 &            204.00 &    153.00 \\
			Max   &            667.00 &              153.00 &            222.00 &    231.00 \\
			\bottomrule
	\end{tabular}}
	\caption{Bare House Low Energy Lighting Upgrade Cost and Savings (12 Lightbulbs).}\label{lightD}
\end{table}

\begin{table}[h]\centering
	{\footnotesize\begin{tabular}{lrrrr}
			\toprule
			{} &  Annual Savings (kWh) &  Annual Savings (KgCO2) &  Annual Savings (GBP) &  Cost (GBP) \\
			\midrule
			Sample Size &            701 &              701 &            701 &    701 \\
			Mean  &          13288.71 &             2170.49 &            185.96 &  11025.41 \\
			Std   &             12.56 &                2.06 &            124.59 &   2003.43 \\
			Min   &          13247.00 &             2164.00 &              1.00 &   5244.10 \\
			25\%   &          13281.00 &             2169.00 &             87.00 &   9589.56 \\
			50\%   &          13289.00 &             2171.00 &            163.00 &  10998.35 \\
			75\%   &          13297.00 &             2172.00 &            263.00 &  12429.73 \\
			Max   &          13336.00 &             2178.00 &            698.00 &  16804.12 \\
			\bottomrule
	\end{tabular}}
	\caption{Bare House Heat Pump Installation Cost and Savings.}\label{hpD}
\end{table}

\begin{table}[h]\centering
	{\footnotesize \begin{tabular}{lrrrr}
			\toprule
			{} &  Annual Savings (kWh) &  Annual Savings (KgCO2) &  Annual Savings (GBP) &  Cost (GBP) \\
			\midrule
			Sample Size &            700 &              700 &            700 &    700 \\
			Mean  &           6962.49 &             1137.22 &             97.27 &  11024.47 \\
			Std   &           1289.74 &              210.65 &             68.06 &   2004.71 \\
			Min   &           3084.00 &              504.00 &              1.00 &   5244.10 \\
			25\%   &           6048.75 &              987.75 &             45.00 &   9589.32 \\
			50\%   &           6932.50 &             1132.00 &             82.00 &  10997.11 \\
			75\%   &           7827.25 &             1278.25 &            136.25 &  12435.23 \\
			Max   &          10609.00 &             1733.00 &            357.00 &  16804.12 \\
			\bottomrule
	\end{tabular}}
	\caption{Cost and Savings of Installing a Heat Pump in a Multi-Project Setup (Double-Glazed Windows + Loft Insulation (15cm)).}\label{hpMD}
\end{table}


\subsection{Dispersion of Savings in the London area}
Here we show the outcome of the formulas developed in Section \ref{sec2} applied to all observations (households) present in the London EPC table. We have shown the annual average of energy consumption per flat in Fig.(\ref{fig:i}), in contrast with the average total floor area, per London borough, in Fig.(\ref{fig:ix}). The plots display the strong correlation between energy demand and area, which has been considered in our regression of $E_0$ over House Volume. In Fig.(\ref{fig:ii}) and Fig.(\ref{fig:viii}) we noticed how the demand and area scale in similar fashion compared to flats (i.e. approximately twice), leading to compatible estimates of demand per area. On what follows, we consider the central values of the parameters listed in the previous section and compute relevant statistics from the sample distribution of project cost and annual savings in kWh, $KgCO_2$, GBP. 

\begin{figure}[tbp]
	\centering 
	\includegraphics[width=0.8\textwidth]{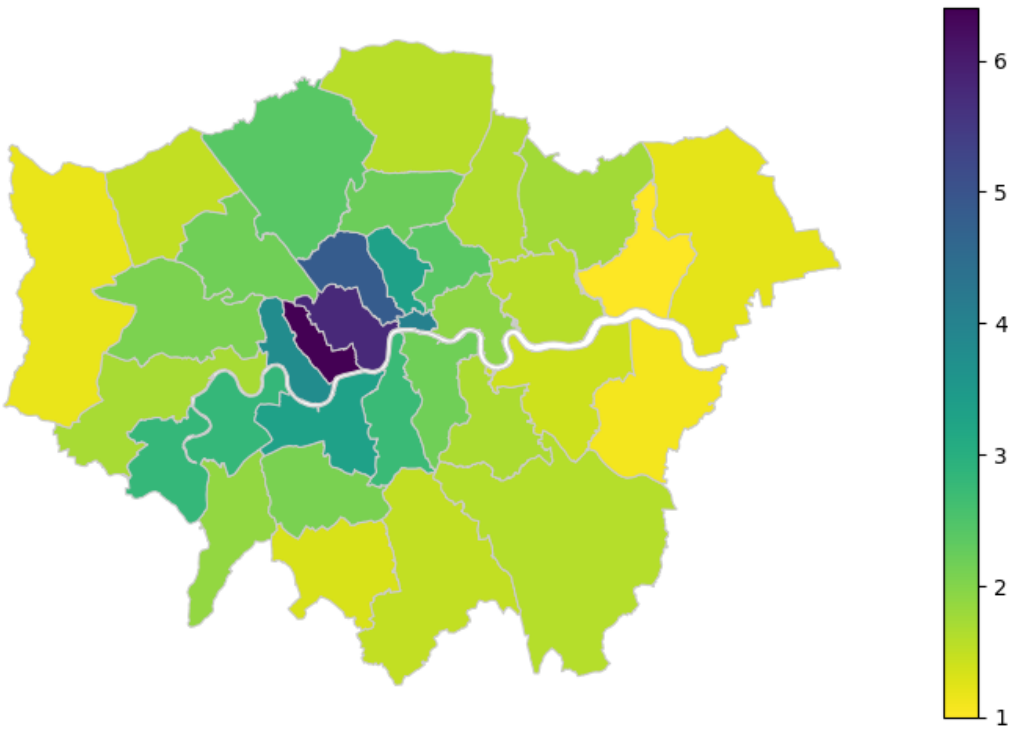}
	\caption{\label{fig:iv} Loft Insulation: Average annual energy savings (MWh) in London Houses, per Borough.}
\end{figure}

\begin{figure}[tbp]
	\centering 
	\includegraphics[width=0.8\textwidth]{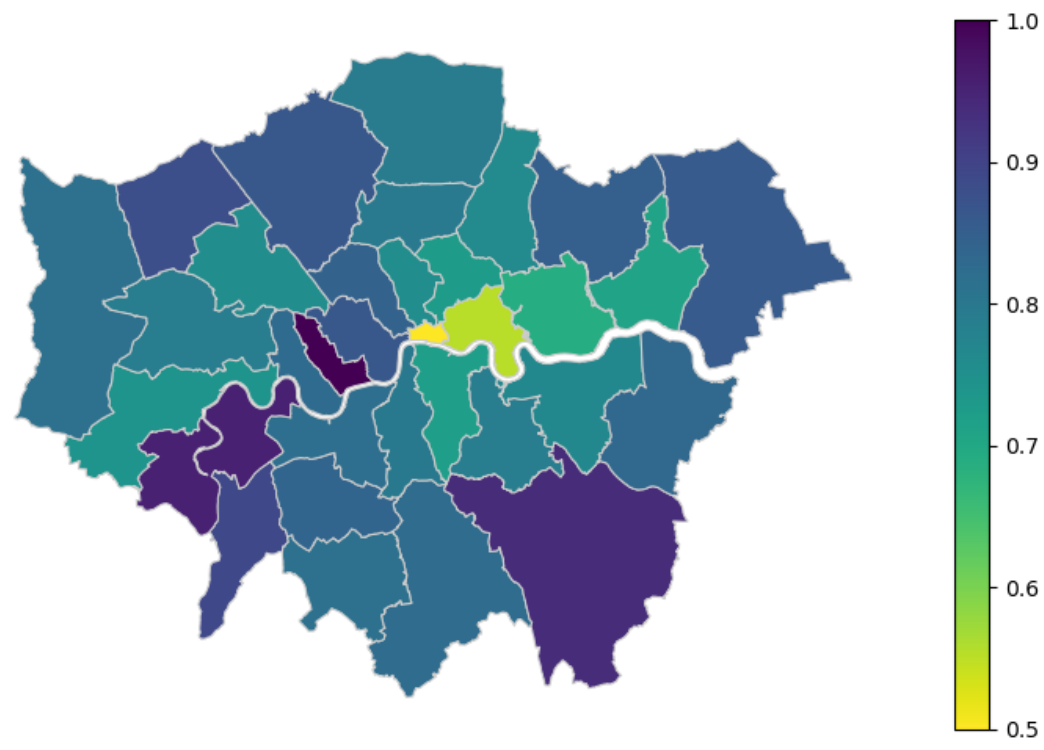}
	\caption{\label{fig:v} Multi-Glazed Windows: Average annual savings (MWh) in London dwellings.}
\end{figure}

\begin{figure}[tbp]
	\centering 
	\includegraphics[width=0.8\textwidth]{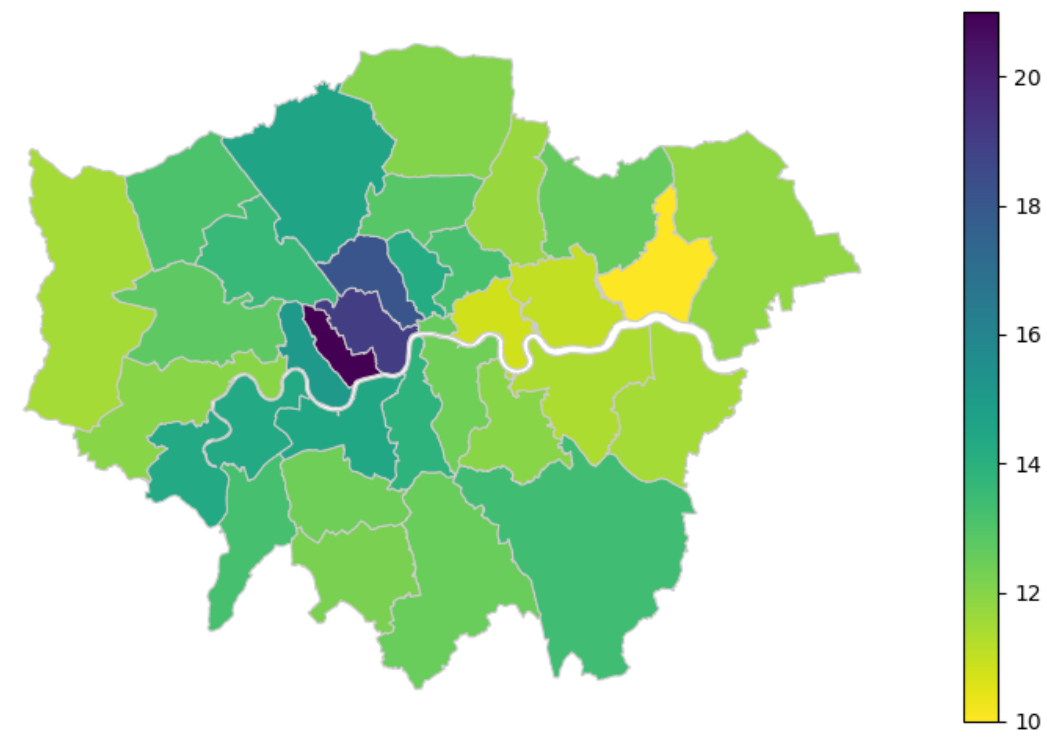}
	\caption{\label{fig:vi} Heat Pumps: Average annual savings (MWh) in London houses.}
\end{figure}

\begin{figure}[tbp]
	\centering 
	\includegraphics[width=0.8\textwidth]{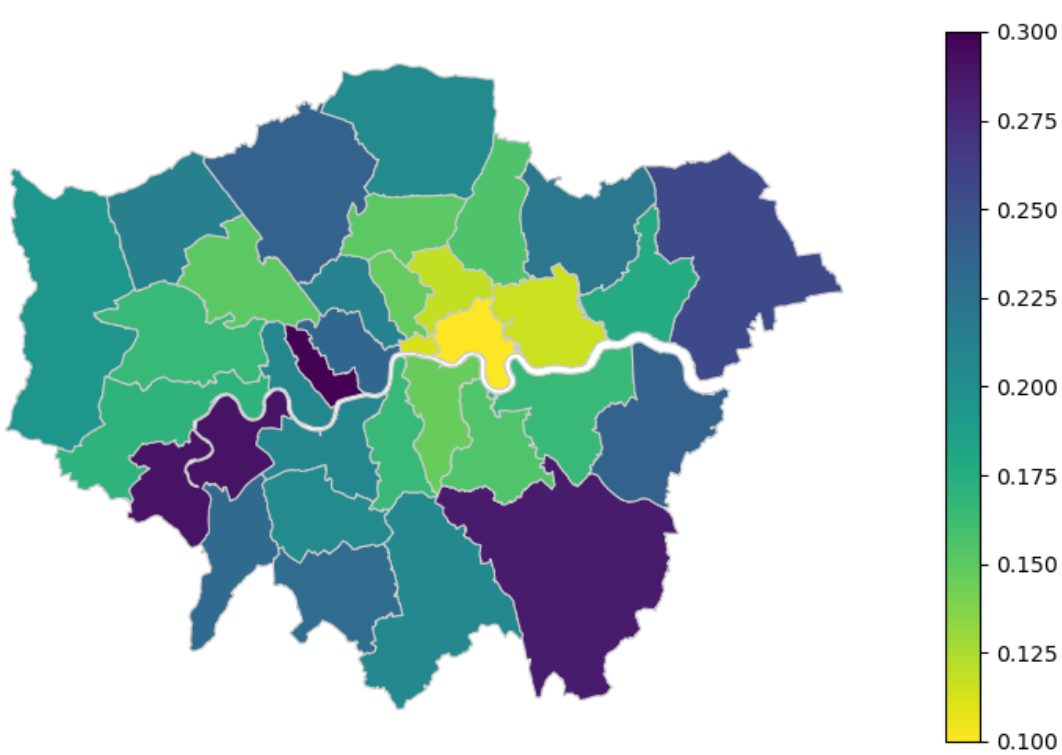}
	\caption{\label{fig:vii} LED Lighting: Average annual savings (MWh) in London dwellings.}
\end{figure}

The level of loft insulation in London buildings has been recorded in the EPC database, and our savings and cost are estimated by suggesting 15cm of insulation material or by supplementing the current level to 15cm. We assume $\Delta T = 10 C$. Similarly, Windows and Lighting profile are measured by EPC variables \textit{Multi-Glaze Proportion} and \textit{Low Ener\-gy Lighting}, respectively, indicating the ratio, in \%, of the presence of these elements in the building. Cost and savings are estimated by filling the ratio up to 100\%. As we discussed in section \ref{e0reg}, on average a dwelling in London will have $85(\pm30)\%$ of its Windows multi-glazed and $53(\pm36)\%$ of its lightbulbs of LED type. Hence, we should expect ave\-rage savings in London to be considerably lower than those predicted for BH. For instance, in Table \ref{Loft} we see that savings are over three times smaller than for BH (Table \ref{insD}). Once more, the discrepancy is not only a consequence of the variety of houses profile, but also from their current degree of insulation. Another relevant information we can extract from Tables \ref{Loft}-\ref{hp}, is the large size of standard deviation, which indicates a flat distribution of savings around the sample mean. In order to identify possible patterns of savings in London's boroughs, we plotted the Heat Maps of Figures \ref{fig:iv}-\ref{fig:vii}. The next section is devoted to our discussion about these and alternative plots. 

\begin{table}[h]\centering
	{\footnotesize\begin{tabular}{lrrrr}
			\toprule
			{} &  Annual Savings (kWh) &  Annual Savings (KgCO2) &  Annual Savings (GBP) &  Cost (GBP) \\
			\midrule
			Sample Size &         810635 &           810635 &         810635 &  810635 \\
			Mean  &           2019.93 &              371.67 &            213.25 &    1009.79 \\
			Std   &           2852.14 &              524.79 &            336.27 &    1054.73 \\
			Min   &              0.00 &                0.00 &              0.00 &       0.00 \\
			25\%   &              0.00 &                0.00 &              0.00 &       0.00 \\
			50\%   &            444.00 &               82.00 &             46.00 &     930.00 \\
			75\%   &           3607.00 &              664.00 &            367.00 &    1480.00 \\
			Max   &         188379.00 &            34662.00 &          29734.00 &   93552.00 \\
			\bottomrule
	\end{tabular}}
	\caption{Loft Insulation: Cost and Savings Summary Statistics (Sample of 810,635 Houses in London).}\label{Loft}
\end{table}

\begin{table}[h]\centering
	{\footnotesize\begin{tabular}{lrrrr}
			\toprule
			{} &  Annual Savings (kWh) &  Annual Savings (KgCO2) &  Annual Savings (GBP) &  Cost (GBP) \\
			\midrule
			Sample Size &        2191227 &          2191227 &        2191227 &  2191227 \\
			Mean  &            801.65 &              147.50 &             94.64 &     2597.13 \\
			Std   &            510.90 &               94.01 &             70.26 &     1460.41 \\
			Min   &              0.00 &                0.00 &              0.00 &        0.00 \\
			25\%   &            512.00 &               94.00 &             58.00 &     1760.00 \\
			50\%   &            731.00 &              135.00 &             82.00 &     2425.00 \\
			75\%   &            996.00 &              183.00 &            113.00 &     3236.00 \\
			Max   &         123554.00 &            22734.00 &          12355.00 &   278495.00 \\
			\bottomrule
	\end{tabular}}
	\caption{Multi-Glazed Windows: Cost and Savings Summary Statistics (Sample of 2,191,227 Dwellings in London).}\label{windows}
\end{table}

\begin{table}[h]\centering
	{\footnotesize
		\begin{tabular}{lrrrr}
			\toprule
			{} &  Annual Savings (kWh) &  Annual Savings (KgCO2) &  Annual Savings (GBP) &  Cost (GBP)  \\
			\midrule
			Sample Size &        2191227 &          2191227 &        2191227 &  2191227 \\
			Mean  &            224.54 &               51.65 &             76.35 &       48.88 \\
			Std   &            267.23 &               61.46 &             90.86 &       63.66 \\
			Min   &              0.00 &                0.00 &              0.00 &        0.00 \\
			25\%   &              0.00 &                0.00 &              0.00 &        0.00 \\
			50\%   &            154.00 &               35.00 &             52.00 &       32.00 \\
			75\%   &            353.00 &               81.00 &            120.00 &       74.00 \\
			Max   &          52490.00 &            12073.00 &          17847.00 &     4988.00 \\
			\bottomrule
	\end{tabular}}
	\caption{LED Lighting: Cost and Savings Summary Statistics (Sample of 2,191,227 Dwellings in London).}\label{light_stat}
\end{table}

\begin{table}[h]\centering
	{\footnotesize\begin{tabular}{lrrrr}
			\toprule
			{} &  Annual Savings (kWh) &  Annual Savings (KgCO2) &  Annual Savings (GBP) &  Cost (GBP)  \\
			\midrule
			Sample Size &         810635 &           810635 &         810635 &  810635 \\
			Mean  &          12630.66 &             2078.41 &            326.56 &   12265.53 \\
			Std   &           5598.23 &              921.64 &            555.50 &    2173.96 \\
			Min   &              0.00 &                0.00 &              0.00 &   11000.00 \\
			25\%   &           9707.00 &             1594.00 &            196.00 &   11000.00 \\
			50\%   &          11489.00 &             1888.00 &            232.00 &   11000.00 \\
			75\%   &          14167.00 &             2332.00 &            290.00 &   16000.00 \\
			Max   &         944821.00 &           154321.00 &          68091.00 &   16000.00 \\
			\bottomrule
	\end{tabular}}
	\caption{Heat Pumps: Cost and Savings Summary Statistics (Sample of 810,635 Houses in London).}\label{hp}
\end{table} 
 
\clearpage

\section{Discussion}
\label{disc}

The work presented a simplified methodology to obtain energy savings estimates resulting from a class of infrastructural upgrades, further applied to the London housing stock and primarily based on the Energy Performance Certificates dataset. Using the expressions described in the Section \ref{sec2}, we calculated the retrofit costs, energy savings, and corresponding $CO_2$ emissions and energy cost reduction, for every property in the London residential sector described by the EPC table. The \textit{projects} considered were the upgrade of loft insulation (houses), the installation of an air-source heat pump (houses), the upgrade of lighting to LED (all dwellings), and the upgrade of windows to double-glazing (all dwellings). The current status of each household was considered, i.e. our estimates for a house whose windows are partly multi-glazed correspond to the outcome of supplementing the remaining single-glazed area. Our main results are the distribution of average annual savings across the London region, grouped by boroughs and represented by the heat maps of Figures \ref{fig:iv}-\ref{fig:vii}. We remember that Loft Insulation and Heat Pumps are house specific projects, and we notice a possible correlation between their outcome with total floor area. Windows and Lighting, on the other hand, define a less specific class and apply to all types of dwellings, in consequence revealing a more uniform pattern of savings across the map. In terms of energy savings, HPs result in an mean outcome around six times the LI value. Dispersion is considerably larger in the latter case, a consequence of the current presence of LI in London houses, in contrast with the uniform absence of HPs , being the main heating system of only 0.15\% of these homes. The LED project energy savings outcome is approximately one fourth of those from multi-glazing. In energy bills, however, the estimates are compatible given the actual difference between gas and electricity costs. 

\begin{figure}[tbp]
	\centering 
	\includegraphics[width=0.49\textwidth]{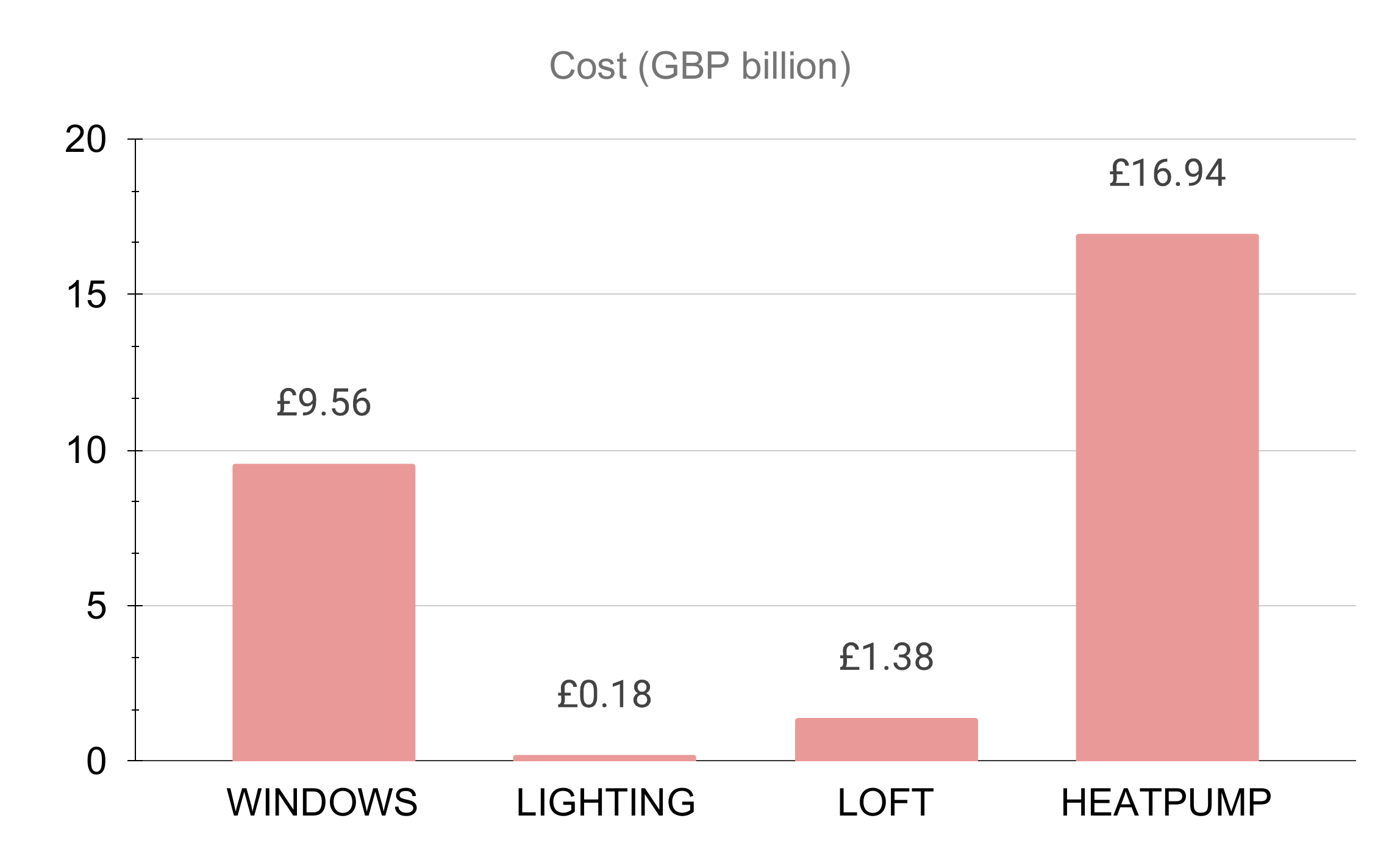}
	\includegraphics[width=0.49\textwidth]{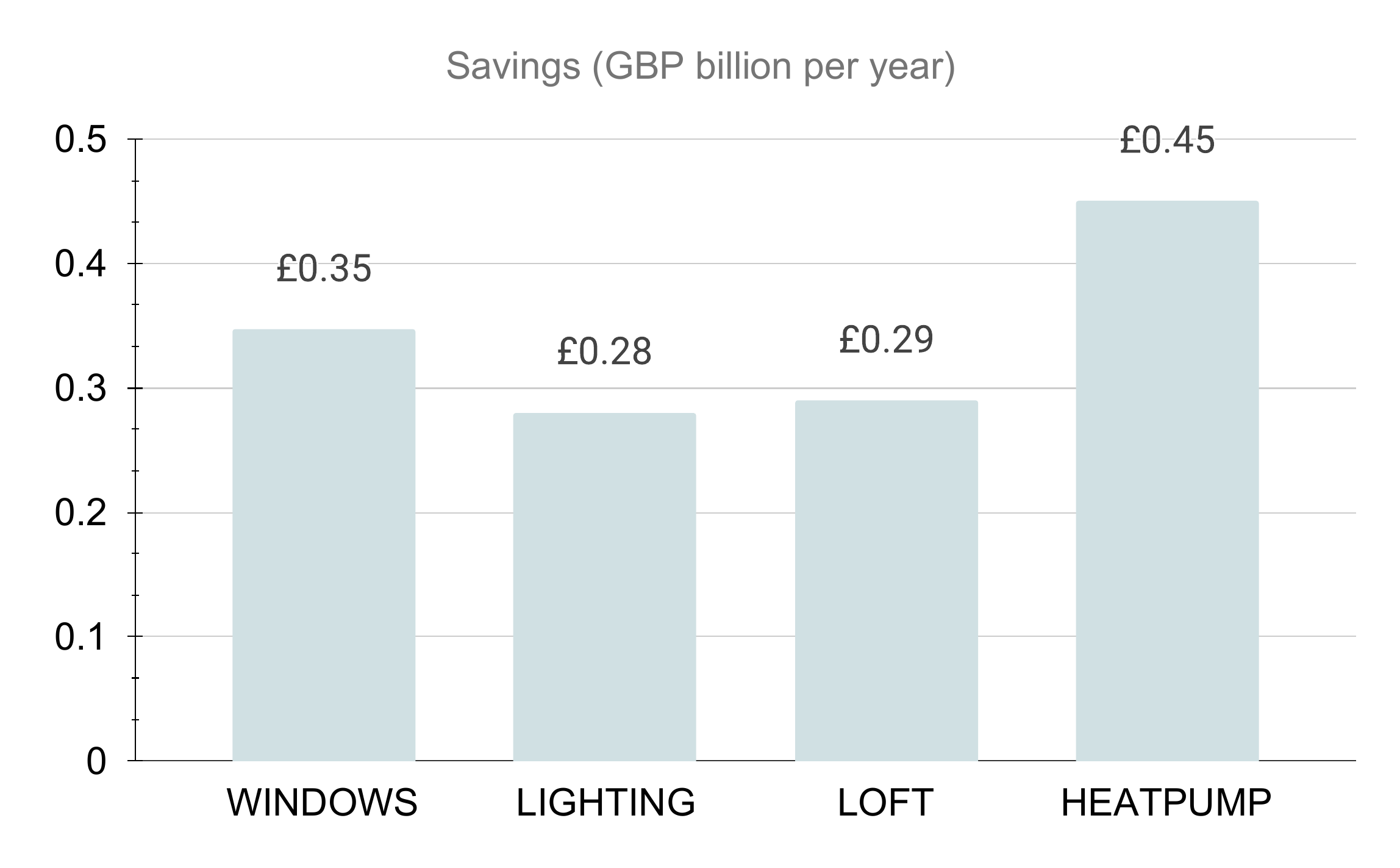}
	\caption{\label{fig:bar} Total cost of four retrofit projects for the entire London area compared to the resulting annual savings, in $\pounds$ billion.}
\end{figure}

We can also use the sample mean of costs and savings in GBP to predict total numbers. In Table \ref{house_ratio} we have the house ratio of London dwellings \cite{dwell}. Applied to the actual number of residences per borough, we can multiply our average predictions and obtain the plots in Figures \ref{fig:hp}-\ref{fig:light}. We notice that, while the persistent location of Barnet and Bromley amid the costly boroughs may be justified by their large areas, the same does not apply to explain Barking and Dagenham or Kingston upon Thames amongst the most economical areas. Taking the total costs of upgrading the entire London housing stock further conclusions can be drawn. Investing $\pounds179(\pm54)$ million to upgrade that stock with LED lighting will bring $\pounds280(\pm9)$ million in savings in the first year, a return on investment (ROI) of eight months. On the opposite side, we have the heat pump approach, which will cost $\pounds17(\pm3)$ billion to install and will lead to savings of $\pounds450(\pm56)$ million per year (more than 37 years ROI). Next, in Eq.(\ref{wind2}) we note that energy savings are maximum if, for $\lambda > 0.6$, we suggest replacing double-glazed with triple-glazed windows, as an alternative to replacing the remaining single-glazed windows with double-glazed ones. As a result, the kWh saved per $\pounds$ invested will drop in boroughs with a high current presence of efficient windows. The result is shown in Figure \ref{fig:ins_win}. In total, multi-glazing windows will cost $\pounds10(\pm2)$ billion, for annual savings of $\pounds347(\pm45)$ million ($\approx$ 28 years ROI). Finally, loft insulation upgrade would cost $\pounds1.4(\pm0.4)$ billion for $\pounds290(\pm36)$ million in annual savings ($\approx$ 5 years ROI). These results are summarized in Figure \ref{fig:bar}. In the energy bill savings bar, we must notice all values at approximately the same size. Despite of leading to the largest amount of energy and CO2 savings, the HP projects switches the heating system from gas to electricity, whose value is currently three times more expensive. We remember from Eq.(\ref{HP3}), that GBP savings are only positive when the cost of gas is greater than the product of HP efficiency times the cost of electricity. Savings from LED and Windows are further scaled from covering a larger spectrum of residences, while the former is particularly impacted by the high cost of electricity. In summary, we can sort the four projects, in terms of their return on investment, by Lighting, Loft, Windows and Heat Pump. Upgrading to LED lighting is a particularly critical investment for the London area.

Potential sources of error in our analysis are related to the presence of outdated cells in the EPC dataset, in addition to our assumptions about energy savings which include the concept of \textit{Bare Home} values. The predicted outcome per project was given in terms of a fraction  of $\alpha_P$ of the BH consumption, whose magnitude was collected either by observing fluctuations around constant mean throughout decades (as in the Lighting case) or by considering approximate absence of project in early years (prior to 1970, in Windows and LI examples) whose data goes back to a period when the quality of retrofit materials can be considered inferior to the modern options. 

We have considered Eq.(\ref{OLS}) as the most illustrative regression model for the estimation of energy consumption by a general household. Home type, built form and age were taken as control variables, shifting the intercept of the demand curve.  In order to adopt a reference energy consumption in a BH, we later re-scaled the estimate by the average presence of `green' elements in London buildings, given by Eq.(\ref{E0}). This simple approach was inspired by the linear relation between thermal energy, temperature and volume. In fact, the conversion between thermal energy and temperature will depend on the \textit{effective} specific heating capacity of the dwelling, which may differ across types and efficiency level. Given that, an alternative version for Eq.(\ref{OLS}) consists of adjusting the slope of the demand curve. We can measure the projects presence either as dummy variable or through a 0 to 1 scale, and regress \begin{equation*}\label{OLS2}
E_0 = (\beta_1 + \sum_j \lambda_j C_j)V
\end{equation*}
where in the set of controls we add the level of Lighting, Windows and Insulation in the observed dwelling. Clearly, here the controls will regulate the slope of the Thermal \-Energy dependence on the Volume, again, as a consequence of the buildings specific heating capacity. Naturally, instead of re-scaling $E_0$, the BH energy demand estimate would be given by resetting the project controls to zero. The analysis of the above model is beyond the scope of this work.

\clearpage
\appendix

\section{Alternative Regression Model and Total Savings Estimate}

\begin{table}[h]\centering
	{\footnotesize\begin{tabular}{lrrrrrr}
			\toprule
			{} &       Coef. &  Std.Err. &       t &  P>|t| &      [0.025 &      0.975] \\
			\midrule
			$\beta_0$        &      761.47 &      4.64 &  164.05 &   0.00 &      752.37 &      770.57 \\
			$Volume$  &        6.77 &      0.01 &  974.48 &   0.00 &        6.76 &        6.79 \\
			$Volume^{-1}$    &    45543.44 &    602.08 &   75.64 &   0.00 &    44363.38 &    46723.51 \\
			$Volume^{-2}$    & -1068552.88 &  19482.42 &  -54.85 &   0.00 & -1106737.77 & -1030368.00 \\
			Flat  &     -287.09 &      1.54 & -186.85 &   0.00 &     -290.10 &     -284.08 \\
			Bungalow  &       65.26 &      4.68 &   13.93 &   0.00 &       56.08 &       74.44 \\
			Maisonette  &     -227.39 &      2.38 &  -95.58 &   0.00 &     -232.05 &     -222.72 \\
			Park Home  &      147.62 &    180.02 &    0.82 &   0.41 &     -205.21 &      500.46 \\
			Semi-Detached &     -155.42 &      2.15 &  -72.32 &   0.00 &     -159.63 &     -151.21 \\
			End-Terrace &     -202.92 &      2.21 &  -91.70 &   0.00 &     -207.26 &     -198.59 \\
			Mid-Terrace &     -403.30 &      2.00 & -201.69 &   0.00 &     -407.22 &     -399.38 \\
			Enclosed End-Terrace &     -230.76 &      3.84 &  -60.14 &   0.00 &     -238.28 &     -223.24 \\
			Enclosed Mid-Terrace &     -343.73 &      3.89 &  -88.38 &   0.00 &     -351.35 &     -336.11 \\
			1900-1929   &      -69.70 &      1.92 &  -36.29 &   0.00 &      -73.47 &      -65.94 \\
			1930-1949   &     -159.23 &      2.01 &  -79.15 &   0.00 &     -163.17 &     -155.29 \\
			1950-1966   &     -299.07 &      2.25 & -132.72 &   0.00 &     -303.49 &     -294.66 \\
			1967-1975   &     -311.54 &      2.46 & -126.55 &   0.00 &     -316.37 &     -306.72 \\
			1976-1982   &     -486.04 &      3.15 & -154.44 &   0.00 &     -492.20 &     -479.87 \\
			1983-1990   &     -475.26 &      3.02 & -157.43 &   0.00 &     -481.18 &     -469.34 \\
			1991-1995   &     -494.55 &      3.59 & -137.73 &   0.00 &     -501.59 &     -487.51 \\
			1996-2002   &     -666.21 &      3.14 & -212.46 &   0.00 &     -672.35 &     -660.06 \\
			2003-2006  &     -801.02 &      3.31 & -242.02 &   0.00 &     -807.50 &     -794.53 \\
			2007-2011  &     -913.13 &      3.49 & -261.44 &   0.00 &     -919.98 &     -906.29 \\
			2012-2022  &    -1397.86 &      4.36 & -320.63 &   0.00 &    -1406.40 &    -1389.31 \\
			\bottomrule
	\end{tabular}}\caption{OLS regression results for the van der Walls model.}\label{tab2}
\end{table}

\begin{table}[h]\centering
	{\tiny}
	\begin{tabular}{lr}
		\toprule
		London Borough &  House/Dwelling ratio \\
		\midrule
		Barking and Dagenham &         0.62 \\
		Barnet &         0.42 \\
		Bexley &         0.61 \\
		Brent &         0.33 \\
		Bromley &         0.56 \\
		Camden &         0.10 \\
		City of London &         0.01 \\
		Croydon &         0.50 \\
		Ealing &         0.41 \\
		Enfield &         0.50 \\
		Greenwich &         0.45 \\
		Hackney &         0.16 \\
		Hammersmith and Fulham &         0.21 \\
		Haringey &         0.32 \\
		Harrow &         0.54 \\
		Havering &         0.61 \\
		Hillingdon &         0.59 \\
		Hounslow &         0.46 \\
		Islington &         0.13 \\
		Kensington and Chelsea &         0.13 \\
		Kingston upon Thames &         0.50 \\
		Lambeth &         0.20 \\
		Lewisham &         0.35 \\
		Merton &         0.51 \\
		Newham &         0.45 \\
		Redbridge &         0.55 \\
		Richmond upon Thames &         0.49 \\
		Southwark &         0.21 \\
		Sutton &         0.52 \\
		Tower Hamlets &         0.11 \\
		Waltham Forest &         0.46 \\
		Wandsworth &         0.26 \\
		Westminster &         0.08 \\
		\bottomrule
	\end{tabular}
	\caption{Ratio of Houses over total number of Dwellings, per London Borough \cite{dwell}.}\label{house_ratio}
\end{table}

\begin{figure}[tbp]
	\centering 
	\includegraphics[width=1.\textwidth]{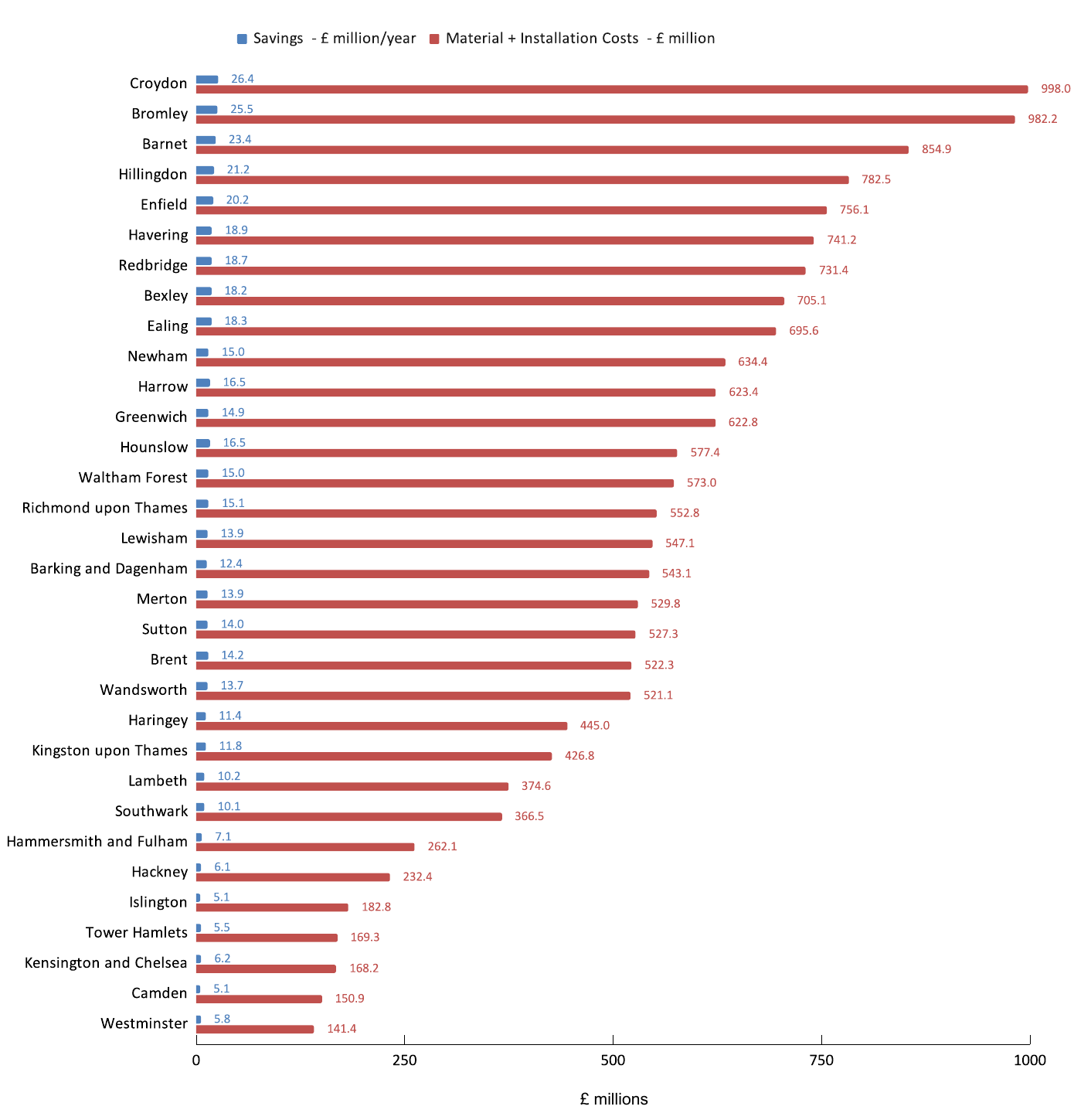}
	\caption{HEAT PUMP: Total installation plus material costs and annual savings (\pounds million), considering all Houses, per London Borough.}\label{fig:hp} 
\end{figure}

\begin{figure}[tbp]
	\centering 
	\includegraphics[width=1.\textwidth]{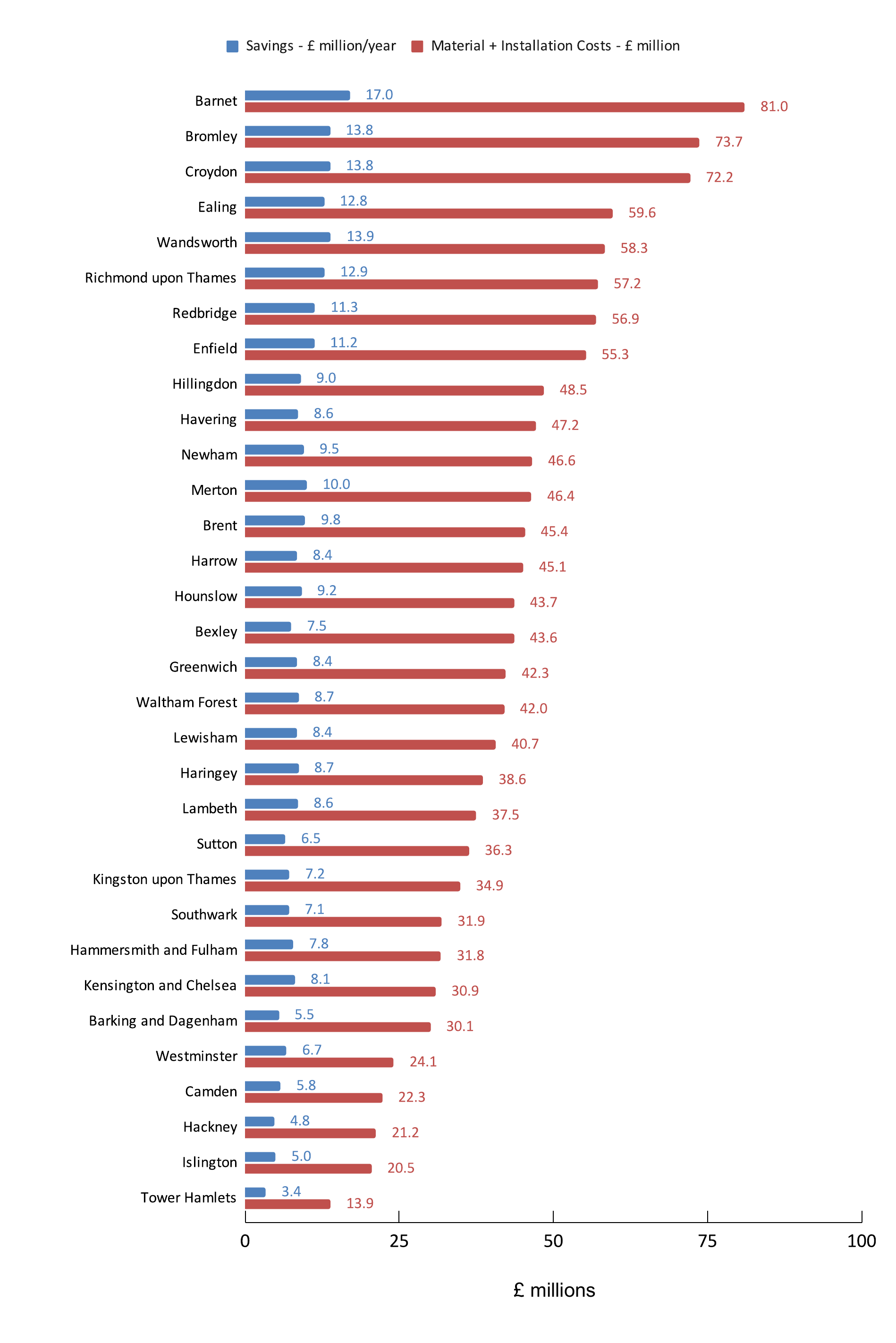}
	\caption{LOFT INSULATION: Total installation plus material costs and annual savings (\pounds million), considering all Houses, per London Borough.}\label{fig:ins_abs}
\end{figure}

\begin{figure}[tbp]
	\centering 
	\includegraphics[width=1\textwidth]{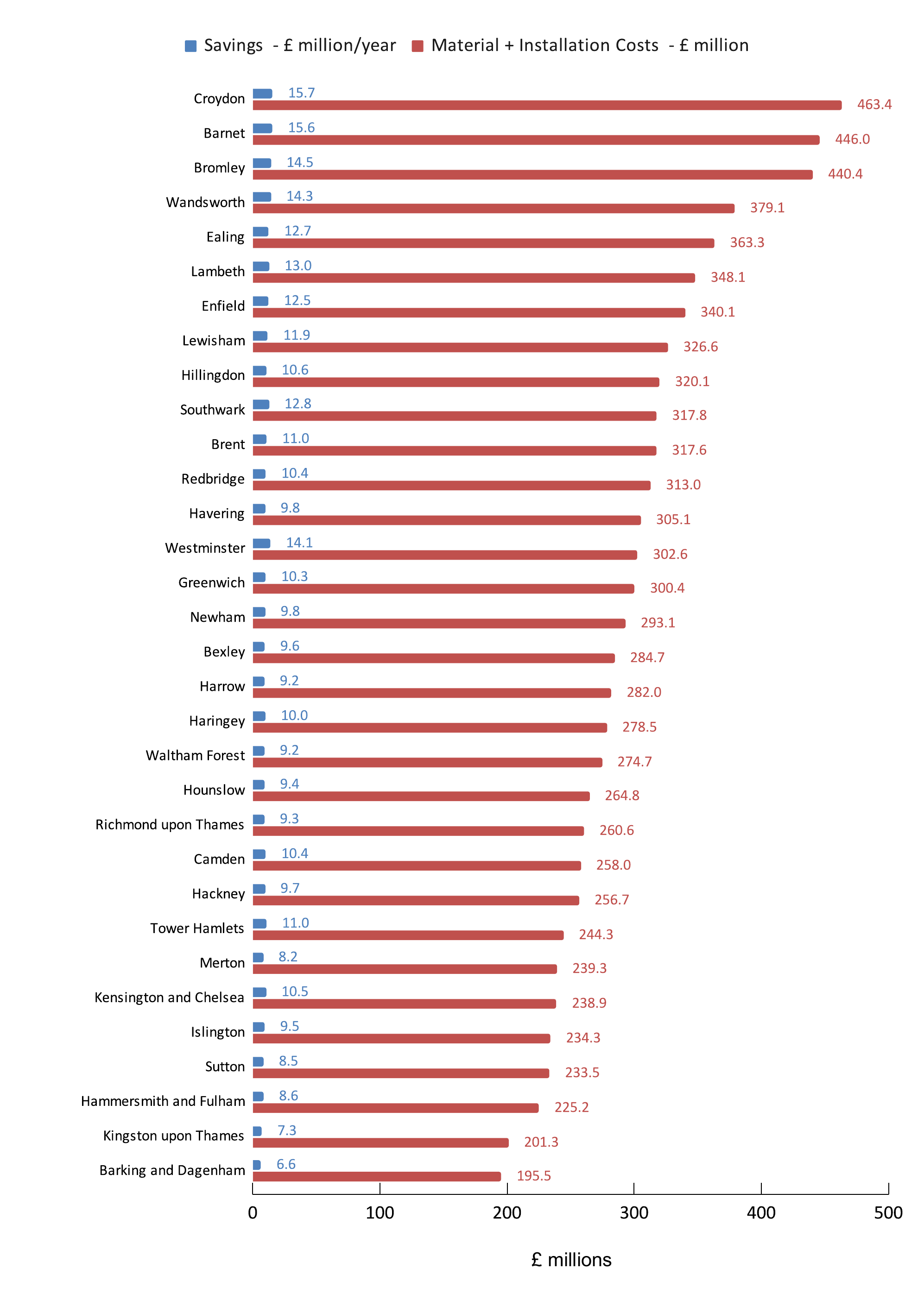}
	\caption{MULTI-GLAZING WINDOWS: Total installation plus material costs and annual savings (\pounds million), considering all Dwellings, per London Borough.}\label{fig:ins_win}
\end{figure}

\begin{figure}[tbp]
	\centering 
	\includegraphics[width=1\textwidth]{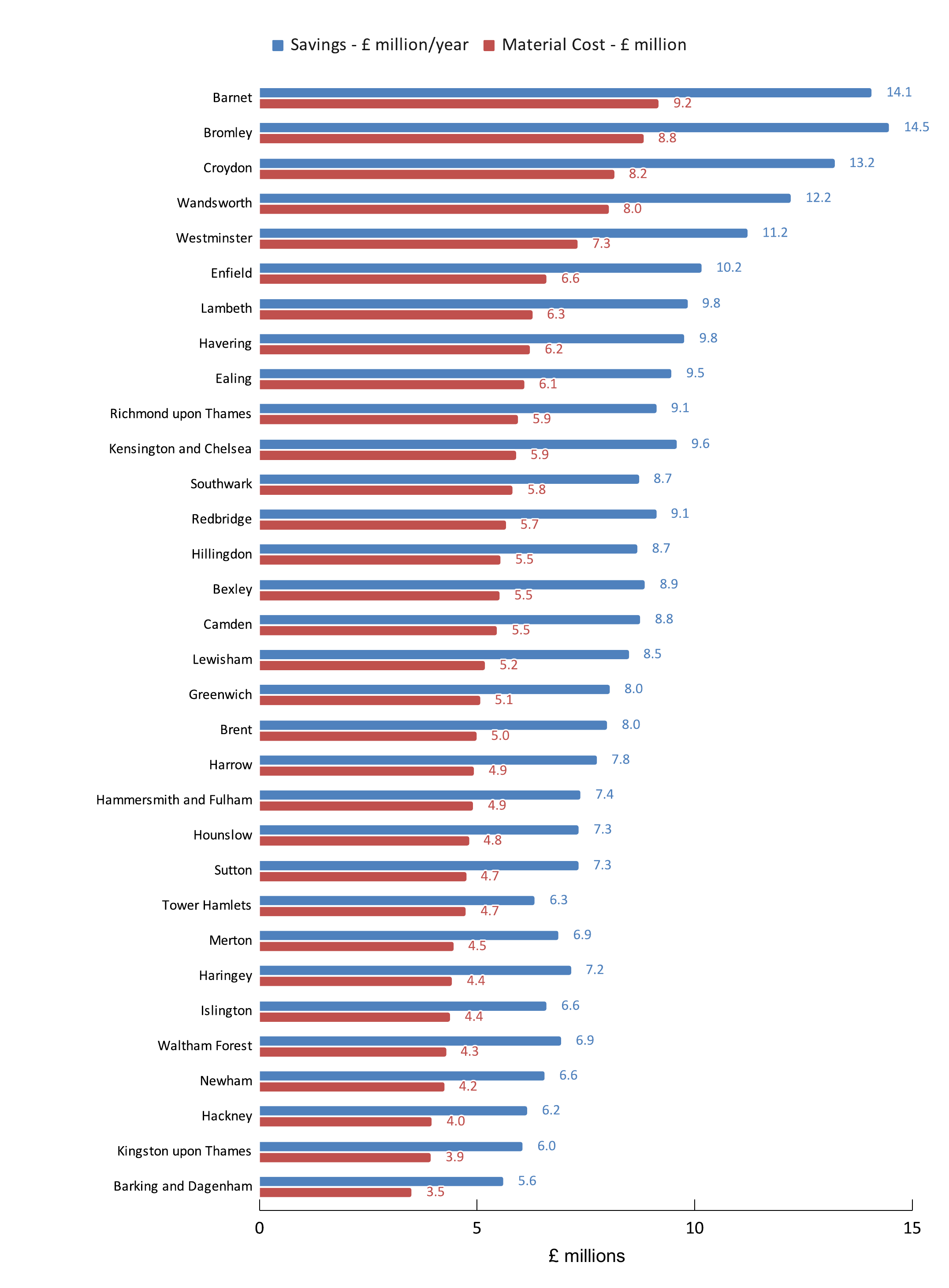}
	\caption{LOW ENERGY LIGHTING - LED: Total cost of installation and annual savings (\pounds million), considering all Dwellings, per London Borough.}\label{fig:light}
\end{figure}

\clearpage
\acknowledgments
Special thanks to Bryan Charter, Rosie Morgan, Moacyr Franco, Eduardo Antoniazzi, Will Meister and Leanne Woodruff for their collaboration, important discussions and inspiration as members of the SuSy.house team.

\bibliographystyle{JHEP}
\bibliography{biblio}









\end{document}